\documentstyle[preprint,prc,aps,epsfig]{revtex}  
\def\shiftleft#1{#1\llap{#1\hskip 0.04em}} 
\def\shiftdown#1{#1\llap{\lower.04ex\hbox{#1}}} 
\def\thick#1{\shiftdown{\shiftleft{#1}}} 
\def\b#1{\thick{\hbox{$#1$}}} 
\begin{document} 
\title{The new mechanism for intermediate- and short-range nucleon-nucleon  
interaction} 
 
\author{Amand Faessler} 
\address{Institute for Theoretical Physics, 
University of T\"ubingen, Auf der Morgenstelle 14,\\ 
D-72076 T\"{u}bingen, Germany}

\author{Vladimir I. Kukulin, I. T. Obukhovsky and V. N. Pomerantsev, } 
\address{Institute of Nuclear Physics, Moscow State 
University, 119899 Moscow, Russia} 
\maketitle 
 
\begin{abstract} 
Arguments against the traditional Yukawa-type approach to $NN$  
intermediate- and short-range interaction due to scalar-isoscalar meson
exchange are  presented. Instead of the Yukawa mechanism for
intermediate-range attraction  some new approach based on formation of the
symmetric six-quark bag in the state   $|(0s)^6[6]_X,L=0\rangle$ dressed due
to strong coupling to $\pi$, $\sigma$ and  $\rho$ fields are suggested. These
new mechanism offers a strong
intermediate-range attraction which replaces the  effective  
$\sigma$-exchange (or excitation of two isobars in the intermediate state)
in   traditional force models.  A similar mechanism with vector
$\rho$-meson production in the intermediate six-quark state is expected to 
lead to
a strong short-range spin-orbital nonlocal interaction in the $NN$  system,
which may resolve the long-standing puzzle of the spin-orbit force in baryons   
and in two-baryon systems.  Illustrative examples are developed which
demonstrate clearly how well the  suggested new model can reproduce $NN$ data.
Strong 
interrelations have been shown to exist  between the proposed microscopic 
model and the one-component Moscow  $NN$ potential developed by the authors
previously and also with some hybrid   models and the one-term separable
Tabakin potential.   The new implications of the proposed model for  nuclear
physics are discussed. 
\end{abstract} 
 
\section{Introduction} 
Since the middle of thirties when Yukawa proposed~\cite{yuk35} his 
classic theory of the nuclear force, based on meson
exchange  between nucleons this concept, although improved and partially also
modified during the last half century (see 
e.g.~\cite{breit60,breithul60,Nijm,Mach87,Mach89,Lac81} and some review of
works  done until  1978 in the book of Brown and Jackson~\cite{Bra78}), is
basically still  the same: the nuclear
force is  assumed to originate from the exchange of one or a few mesons
between {\em  isolated} nucleons. Though in the last two decades one added to
the nucleons also other channels with one or two $\Delta$-isobars in the  
intermediate  state~\cite{Mach87,Mach89}, with isobars again interacting via 
meson exchange. 

Based on this concept a large variety of potential models have been 
suggested in   recent years to describe the  $NN$ interaction, which fitted
very accurately the  experimental data for $NN$ scattering until the energy
300 MeV in the  laboratory  system. 

However with the accumulation of many new data in the field of hadronic
physics  it became more and more evident that the traditional $NN$
interaction models  (i.e. based on the meson exchange concept) suffers from
numerous inner  inconsistencies and discrepancies when e.g. the same
meson-nucleon form factors  have to have a different short-range behavior
while describing very similar  processes. In particularly, the same functional
form of the $\pi NN$
form factor $F_{\pi NN}(q^2)$ has  to have very different cut-off
parameters $\Lambda_{\pi NN}$ to describe  elastic and inelastic $NN$
scattering, or in description of two-body $2N$ and  three-body $3N$ forces
etc. (Some other numerous examples of such inconsistencies  are discussed in
Sect.II). 

At the same time, due to radical improvements of the accuracy and the reliability of 
dynamical few-nucleon calculations, one begins to find also some numerous 
disagreements between the new experimental data and the results of the most 
accurate Faddeev calculations (for a list, although far from complete, of
such  disagreements in few-nucleon calculations see e.g. in~\cite{Kuk99}). It
is very  instructive that many of such disagreements cannot be removed by
introducing  phenomenological $3N$ forces into the 
calculations~\cite{Kuk99,Pick92,Torn98,FB98}. 

Some recent works in the field based on chiral perturbation theory 
($\chi$PT) may serve as  a very clear indicator for the degree of our
understanding (or misunderstanding)  of the  fundamental $NN$ interaction.
This is especially true for the   works~\cite{Kai97,Kai98}. There the authors
have  shown that within chiral perturbation theory without introduction
of any  cut-offs it was impossible to describe all the lowest partial waves
even if   one incorporates the excitation into intermediate $\Delta$-isobars
and the vector   (viz. $\rho-$ and $\omega$-) meson exchanges. Thus, the
quantitative description  of lowest partial waves with $L=0 - 2$ up to 
$E_{\rm lab}=300$~MeV  requires already to go beyond the framework of the $\chi$PT.
This problem becomes  more urgent in passing to the intermediate energy
region around $E_{\rm  c.m.}\simeq 1$~GeV where a strong coupling to the
meson production channels will  require that the application of $\chi$PT gets
even more complicated. 

On the other hand, we consider critically in the Section~III the problem of
the existence and the role played in the fundamental $NN$ force by a 
scalar-isoscalar light meson usually referred to as $\sigma$-meson.  The
$\sigma$-meson exchange is considered in the traditional OBE models as a main
contribution responsible for the strong intermediate-range attraction between
nucleons and eventually as the  main component of nuclear binding (e.g. in the
Walecka model). Despite of the very numerous attempts to find a well
developed resonance in the $\pi \pi$ $s$-wave system undertaken in recent
years, no definite evidence for such a well defined resonance has been
found (see e.g. the recent review~\cite{Span98}). It is likely there is no
such light scalar meson in free space. 

Moreover,very recent studies of different groups have 
demonstrated~\cite{Kai97,Kai98,Oset99} that the exchange of a correlated
$\pi\pi$ pair in an $S$-state between nucleons leads to a {\em repulsive} rather
than {\em attractive} contribution in the $NN$ interaction. Thus, we
should attribute the $NN$ intermediate-range attraction to a generation
of the two intermediate $\Delta$-isobars (or at least to an $N\Delta$
intermediate  state)~\cite{Kai97,Kai98}. But as it will be argued in
Sect.~IV, this intermediate $\Delta\Delta$ state has a strong overlap to
the symmetric six-quark state $|(0s)^6[6]_X,L=0,2>$ and thus the above
$\Delta\Delta$ state can be replaced by an intermediate symmetric
six-quark state strongly coupled to the $2\pi$-channel. 

Thus  we tried to circumvent the problem in the treatment of lower partial waves
by  refraining from the basic Yukawa idea of the meson exchange between 
(isolated) nucleons and we develop some new interaction mechanism on the
basis  of a quark model where quarks are strongly coupled to chiral fields.

Our treatment is based essentially on the group-theoretical considerations  
of the symmetries and on  the algebraic recouplings in the six-quark   system
and the specific role played by fully the symmetric six-quark
state  $|(0s)^6[6]_x[f]_{cs}\rangle$ in the  $NN$ interaction in lower partial
waves.   In particularly, one could even expect that such a fully symmetric
$6q$ state, due to  the  maximal overlap of all six quarks (which implies some
enhancement of  $q\bar{q}$ fluctuations inside such a state), may lead in the
direction of a  phase transition of the chiral symmetry (or partial chiral
symmetry) restoration. This Goldstone limit, or even only approaching
this limit, means,  in accordance with the variational principle,
the appearance of a strong additional  attraction between quarks and thus 
also between two 
nucleons at intermediate  range (i.e. at distances $r_{NN} \sim
0.7 \div 1.2$~fm where such a dressed six-quark  bag is localized). 

The physical assumption about such a possible phase transition in a fully 
symmetric multiquark state or, to be more correct, about the approach to such
a limit with an increasing density of quarks is the basis of the 
new model for  the $NN$ interaction presented here. Its rigorous justification can be 
found only by careful studies of chiral dynamics of the multiquark system with 
a detailed treatment of the $qq$ correlations and strong coupling of such 
symmetric multiquark states with the Dirac sea of antiquarks. This problem 
which combines the chiral field theory and relativistic multiquark dynamics 
is a too complicated many-body problem to be handled today. However we do believe that
it   is already now possible to construct a chiral quark model with a   minimal
number of adjustable parameters, which will be able to describe   the $NN$ 
interaction in both lower and higher partial waves. To describe the latter
our  model should be combined with the presently already well developed   
$\chi$PT approach with $\pi$- and $\pi\pi$-exchanges between two nucleons. 

How well the proposed model may work is illustrated by a 
simple model suggested here on the basis of the proposed new mechanism in 
Sect.V. In particular, the above simple model can describe perfectly all the 
lower $NN$ phase shifts in a rather large energy range $0 \div 600$~MeV. Hence, 
though we are unable presently to justify strictly the suggested new model,  
its general framework looks completely natural and is in accordance with   
the general concepts of quark models and chiral field dynamics. 
 
The organization of this paper is following.  In Sec.~II we offer a critical look
to the OBE models and discuss the  difficulties of traditional meson-exchange
models with anomalously high cut-off  parameters $\Lambda$ and also with
respect to their application in few-nucleon  problems. Section~III includes a
critical discussion of the  scalar meson puzzle in  the light of some new
results. In particularly, we argue here that  the exchange by a $S$-wave
$\pi\pi$ correlated pair with no $\Delta$-isobar in  the intermediate state
does not lead to any significant intermediate-range  attraction (which one
ascribes usually to the $\sigma$-meson exchange) but  rather results in a
{\em repulsion} between two nucleons. In Sec.~IV we describe in  detail
the new model for intermediate and short-range interaction and compare  it
with the  traditional Yukawa mechanism of $\sigma$- and $\rho$-exchange.
Section~V is devoted to interrelations between the $NN$ interaction model
suggested   in this   work and the potential models proposed previously. In
particular,   we compare the new approach with the Moscow $NN$
potential developed  in previous years, with the hybrid quark compound bag
(QCB) model   and with the one-term separable Tabakin potential. We 
elucidate the microscopic basis for the above models. In Conclusion we 
summarize the main results of the work. Some algebraic details required 
for the derivation of the  basic formulas and some tables of the
group-theoretical algebraic coefficients are presented in the Appendix.

\section{Critique of the  basic assumptions of OBE models} 
Despite of the  relative success in the  description of  the low-energy $NN$ 
scattering data up to $E_{\rm  lab}=350$~MeV, the traditional OBE models
based on  the initial  Yukawa  meson-exchange mechanism for the
nucleon-nucleon force are  suffering from many inner contradictions and
inconsistencies. These contradictions concern not only the description of
the  $NN$ data themselves but  also e.g. the description for few-body data. 
All these
contradictions and inner  inconsistencies form a large array of discrepancies
either with accurate  experiments or other existing theories and these
drawbacks seem   to be almost unremovable today  because an improvement in
one point leads very often to  an appearance of discrepancies in other
places. To avoid   repetitions we arrange the discussion according to the
following points. 

\subsection {The range of  the $NN$ force due to heavy meson exchange and the quark radius  
of the nucleon.} 
 
While the range of  the $\pi$-exchange force (OPE) $\lambda_{\pi}\simeq 1.45$~fm
is  much larger than the quark radius of the nucleon $ <r_N> \simeq 0.6$~fm
so that   the  Yukawa $\pi$-exchange may be considered to occur mainly
between two separated  nucleons, the heavy-meson exchange with masses $m\ge
600$~MeV occurs mainly on  the distances $r_m \simeq 0.2\div 0.8$~fm where
the  two nucleons are   strongly overlapping.  Thus this heavy meson exchange
happens mainly in the field of all six  quarks of the  participating
nucleons. Hence in OBE models using such a heavy-meson  mechanism it is first
necessary to justify the employment of "free-space"  meson-nucleon coupling
constants and cut-off form factors. As a result of this,  all existing OBE
models have severe problems with the  short-range cut-off  parameters
$\Lambda$~\cite{Mach89,Lac81,Kuk99,Pla94,Kondr90,Uz98} (see especially the severe 
critique in~\cite{Pla94} and also in the next  Subsection). Thus all the
short-range parts of OBE potentials are treated  purely
phenomenologically~\cite{Mach87,Mach89,Lac81,Bra78} but using at the same time
the Yukawa framework which   looks rather inadequate for such short ranges.
Very recently  an attempt~\cite{Kai97,Kai98} undertaken to refrain from this
short-range   phenomenology but staying   still within the framework of a
meson-exchange model (with a perturbative  chiral field-theory treatment of
two-pion exchange) has demonstrated     very  clearly that the OBE+TPE model
even taken consistently (and without  phenomenological cut-off form factors)
is able to describe only the higher $NN$  partial waves. Hence the
description of lower partial waves demands a  non-perturbative dynamical
treatment. 

\subsection{The difficulties with short-range cut-off form factors} 

It is well known~\cite{Mach87,Mach89,Pla94} that in all the  OBE models 
the values of the    cut-off
parameter   $\Lambda_{mNN}\;(m=\pi ,\sigma ,\rho , \omega ...)$ are strongly
increased as  compared with any microscopic model for meson-baryon coupling
and also as  compared with fits to the data of meson-nucleon scattering 
experiments~\cite{Lu95,Bo99,Kondr90,Uz98}.  This disagreement is especially
evident in the values of $\Lambda_{\pi NN}$  which can be derived from the
theory of $\pi NN$ form  factors~\cite{Pla94,Lu95,Eric} and  even from direct
experiments $N(e,e'\pi)N'$ in which a pion is knocked out   from the pion
cloud of the nucleon by fast electrons~\cite{Yudin}. 

In any case the values of $\Lambda_{\pi NN}$ taken in all OBE models to  
fit the $NN$ data lie in the interval~\cite{Mach89,Pla94}: 
$$  
\Lambda_{\pi NN}^{\rm OBE}\simeq 1.3 \div 2.0 \mbox{ GeV} 
$$ 
while all above mentioned direct estimates and experiments result in the values: 
$$ 
\Lambda_{\pi NN}^{\pi N + \rm theor} \simeq 0.4 \div 0.8 \mbox{ GeV}, 
$$ 
i.e. a discrepancy of a factor 1/3 to 1/4 or even less. 
 
Moreover, the choice of the strongly increased values   $\Lambda_{\pi NN}
\simeq 1.3 \div 2.0$~GeV in microscopic nuclear models results  in a strong
enhancement of the pion field inside nuclei~\cite{Sliv85} which is in   
drastic  disagreement with many observations (see the numerous examples in the
review~\cite{Sliv85}). Thus, if even one assumes these strongly enhanced values
of  $\Lambda_{\pi NN}^{\rm OBE}$ required for  the $NN$ interaction  in OBE  
models as some {\rm effective description}   of an unknown short-range part
of the  $NN$-interaction, this assumption   turns out to be unacceptable for
the  description of the pion dynamics   in nuclei with such models. But this
is not the end of the story. 

Even if one forgets for the moment the nuclear pion dynamics, the large value of  
$\Lambda_{\pi NN} \simeq 1.3 \div 2.0$~GeV seems to  be fully incompatible  
with the description of pion production in collisions $pp \to
pn\pi^+$~\cite{Lee84} and also  with elastic backward $p+d$
scattering~\cite{Kondr90,Uz98}. Let us to add to this collection still an 
another example: a small value of  $\Lambda_{\pi NN} = 0.528$~GeV has been
found by T.~Cohen~\cite{Coh86} in his analysis for $\pi NN$ form factor
within the Skyrme model. A very similar value  $\Lambda_{\pi NN} =0.63$~GeV
(and $\Lambda_{\rho NN} =0.7$~GeV) has been extracted~\cite{Dmitr86} from an analysis of
exclusive experiments $NN \to N\Delta$ at incident proton energies a few GeV.
There are also many other evidences which point very unambiguously to the
necessity for soft cut-off parameters  $\Lambda_{\pi NN}$ and  $\Lambda_{\rho
NN}$ for both the $\pi NN$ and $\rho NN$ form factors.    Last not least the
$3N$-force models (via  pion-exchanges) which describe accurately the  $3N$-
and  $4N$-systems~\cite{Gibs86,Glo96,WirFB97} need still a soft cut-off
parameter   $\Lambda_{\pi NN}$. 

Quite a similar situation is observed also for other mesons, like $\sigma$, 
$\rho$ and $\omega$ for which one needs also large cut-off parameters  
$\Lambda$ in OBE models as compared to values given e.g. by the
vector-dominance model  (in case of $\rho$-mesons).  In total, the problem
with artificially enhanced values of the cut-off parameters seems to be almost
unavoidable in the OBE models. For example, in the attempts to solve this 
problem
Ueda~\cite{Ueda91} suggested to add the three-pion exchange contributions in
the form of $\pi\rho$ and $\pi\sigma$ terms and also some "heavy" pion $\Pi$
exchange. He found again that the cut-off parameter $\Lambda_{\Pi NN}$ for
the $\Pi$-meson should be about 3~GeV (!) to fit the $NN$ scattering data.

Therefore in all these cases, i.e. really for the description of whole
short-range  part of the $NN$ interaction the Yukawa model shows an   inner
inconsistency even if  the form factors are considered as an {\em effective} 
description of the  interaction. A very similar critique of the  short-range
part of  the $NN$   interaction in the current OBE models has been presented
also by the Bochum   group~\cite{Pla94}.   In the next Section we will
demonstrate that there are also serious problems  in a consistent
interpretation of OBE models at intermediate ranges.  

\subsection{Few-body puzzles originating from the  application of the conventional 
$NN$ interaction models to precise few-nucleon calculations} 
 
In recent years in high precision few-nucleon  calculations which  
use the most
realistic conventional $NN$   potentials  for low ($<200$~MeV) and
intermediate energies (200 -- 300~MeV),  numerous marked disagreements with
accurate modern experimental  data have been 
found~\cite{FB98,Glo96,Wit98,Koike97,Fons98,Wit91}.   Because the full list
of these disagreements and puzzles is rather long~\cite{Kuk99} we will
present here only the newest or best known ones. 
\begin{itemize} 
\item[(i)] The best known disagreements have been found since the middle
of  seventies in $3N$- and $4N$-binding energies. The strong underbinding
found in  the $3N$ and $4N$ ground-state energies have been explained long
ago with a   significant  contribution of a meson-exchange $3N$
force~\cite{Gibs86,Glo96}. However this $3N$ force did  not  help really to
understand 
quantitatively the remaining $3N$ puzzles,  e.g. those
pointed out in (ii) -- (vi) below. Moreover, it was demonstrated very 
recently~\cite{Mey99,Sak99} that the conventional $3N$ forces used  
fail quite evidently in the  treatment of new high-precision 
experiments of $n{\vec d}$ and ${\vec p}d$ elastic scattering at energies
$E_N \simeq 150  \div 300 $~MeV in the backward hemisphere. Thus, the explanation
given with the above $3N$ forces for  binding energy puzzles must be also
considered as an {\em ad hoc} fit to specific data 
(see especially~\cite{Pick92}). 

\item [(ii)] The well known puzzle of the analyzing power $A_y$   for
low-energy ${\vec n}d$ and ${\vec p}d$ scattering~\cite{Wit93}.  In recent years the
situation with this puzzle has not improved but  even got worse. 
The traditional three-nucleon force contribution does not help to remove the 
$A_y$ discrepancy. Moreover, 
very recently we made~\cite{Dol99} a new high precision  calculation for
$A_y$ with our new one-component $NN$ potential (the so called  generalized
orthogonality condition model~\cite{PRC99}), which fits excellently the $NN$
phase shifts in all low partial waves.  We found for the analyzing power  in this calculation almost
the same disagreement with experiment  as for the conventional $NN$ models
like AV18 etc. Because the off-shell behavior  of the above $NN$ model potential
is strongly different from those  for conventional $NN$ models one can
conclude from these results rather reliably  that the explanation of $A_y$
puzzle is not related to different  off-shell behaviors of the  various $NN$
potentials but demands some new type of spin dependent $3N$ forces. 

\item[(iii)] A recent analysis by Scholten et al.~\cite{Schol99} of   new data
from Osaka for $pp \to pp\gamma$ bremsstrahlung at $E_p=390$~MeV
discovered a  large disagreement with predictions of the existing $NN$
potential models. The  data still could be explained by artificial
enhancement of $\Delta$-isobar  current contribution by a factor 1.7. Thus 
the situation here is similar as in   the treatment of short-range $NN$
interaction with conventional force models. 

\item[(iv)] Recently it was found~\cite{Mey99,Sak99,Bod99} that the   so-called
Sagara puzzle (disagreement for the backward $Nd$ elastic  scattering near
the minimum of the cross section) increases with growing energy.   At
$E_N=200$~MeV in the lab system the disagreement is as large as 30\%. 
However if the conventional $3N$ force is taken into account the 
disagreement is considerably  
reduced but instead, there appears some larger disagreement for  the vector
$A_y$   and  the tensor $A_{xx}$ analyzing powers at the same backward
angles. 

\item[(v)] Very significant disagreements were discovered recently in  
${}^3{\rm He}(e,e'p)$ and ${}^3{\rm He}(e,e'pp')$ reactions at moderate to
high  transferred momenta and energies $\omega$~\cite{Gol99,Block99}. In
particular, at energy transfer  $\omega \simeq 200$~MeV one observes in the
first reaction a very big ($\sim150  \div 200 \%$) mismatch between the
complete Faddeev $3N$ calculations and  experimental data. The traditional
MEC contribution does not help. It  modifies the theoretical results
only slightly. 

\item[(vi)] Numerous disagreements with the data have also been found in recent 
four-nucleon calculations of  the Lisbon~\cite{Fons98,Fons99} and   Grenoble
groups~\cite{FB98}. While the theoretical results of both are  in very good 
agreement with each other.   \end{itemize} 

This list may be continued much further (see e.g. recent reviews~\cite{Kuk99}). 
So, the above few-body puzzles and disagreements found very recently together 
with long-standing puzzles are clearly signalling that the existing $NN$-force 
models (based on the meson-exchange mechanism) do not include at 
intermediate and short ranges some important nontrivial contribution.  
A candidate for such a nontrivial contribution which has been fully missed in 
previous $NN$ models is suggested in the present work. 
 
\section{The scalar meson puzzle and the problem of the intermediate-range $NN$ 
force} 
 
The problem with scalar mesons and their role in the  hadron-hadron  
interaction is attracting   increased interest today (see
e.g.~\cite{Span98,Bon99}).   This interest concentrates on the experimental
identification of the   the  scalar mesons and on their contribution to  the
description of hadron collisions, in particular to the  $NN$ interaction. 

According to the traditional point of view advocated for a long time by many 
"constructors" of $NN$ potentials (see e.g.~\cite{Mach87,Mach89}) an exchange
by the  $(\pi\pi)$ correlated pair in relative $s$-wave between two pions
in a  combination with the excitation of intermediate $\Delta$-isobars is
responsible  for the  strong intermediate-range attraction between 
nucleons~\cite{Mach87,Mach89,Durso80}. Further, in  the conventional picture,
this strong attraction at short distances is fully  compensated by a strong
repulsion due to  $\omega$-exchange~\cite{Bra78,Nij93}. 

Very recently, however, it was found by two groups
independently~\cite{Kai98,Oset99} that the $(\pi\pi)$ $s$-wave correlation 
being treated consistently and taken itself,  is unable to give any
intermediate-range attraction but it results even in   a rather  strong
short- and intermediate-range {\em repulsion} between nucleons. Thus, in the
conventional meson-exchange mechanism, the main intermediate-range attraction should
be associated only with the excitation of the intermediate-state
$\Delta$-isobars. Some independent arguments for favour of this 
conclusion follow 
from the obvious failure to get this strong attraction from various
microscopic models like the Skyrme soliton interaction model, the RGM $6q$
treatment with $qq$ interaction based on the Goldstone boson
exchange~\cite{Stancu} in which the $\Delta\Delta$ (or $\Delta N$) state
excitation has been neglected.

A second important argument comes from the experimental search for   the
low-mass scalar-isoscalar meson~\cite{Span98,Bon99}. While the highly
excited scalars  $f_0(1370)$ and $f_0(1500)$ have been identified more or
less reliably in  experiments, the identification of low-mass scalar meson
resonances (which one  refers often to as $\sigma$-meson and which one
relates with the  $\pi\pi$ $s$-wave resonance)  is in no way well accepted. 
The spread in the mass and width  estimates for  these states are
extremely large~\cite{Span98}. The estimates accepted today   are as
follows~\cite{Caso98}: 
$$ m_{\sigma}=400 \div 1200 \mbox { MeV}, 
$$
$$ 
\Gamma _{\sigma} = 300 \div 500 \mbox{ MeV}, 
$$ 
i.e. they are rather uncertain. Thus also from the  experimental side the 
situation looks rather unsatisfactory. Therefore the 
attempt to interpret the basic internucleon attraction as originating from a  
Yukawa-type exchange of a scalar meson (the existence of which as a {\em free 
particle} is doubtful) seems to us not the best way to 
understand the  intermediate-range interaction. 
 
Nevertheless there is no doubt that some scalar meson contribution (of
the  $\sigma$-exchange type) is necessary for understanding
numerous  processes in hadron physics, e.g. for $\pi N$ and $NN$
interactions. Hence the  above deep contradiction should be somehow
resolved. 

We propose here a new approach to solve this puzzle. This approach is in  
part based on the assumption that the scalar-isoscalar  excitation of the QCD
vacuum which is conventionally referred to as $\sigma$-meson  is in essence
not a real particle in  free space (like e.g.  the $\rho$-meson) but some 
sort of quasiparticle excitation {\em inside} hadrons, in particular inside
a  multiquark bag. This quasiparticle can thus exist  inside the six-quark
bag but not in free space. It can be understood very naturally from this
assumption why it was impossible to observe this particle to date in 
the $\pi\pi$ final state interaction. Therefore one can conclude that such an
exchange of a scalar-isoscalar quasiparticle may occur very  naturally in the
field of six valence quarks   but that such a quasiparticle cannot couple  
with isolated nucleons in free space. 

These ideas lead very naturally to  the new basic mechanism of the  
intermediate-range $NN$-interaction presented in following  Section. 

\section{The dressed bag mechanism for the intermediate and short range 
$NN$ force} 
 
In order to give to the reader some clue to the suggested mechanism we  
display the respective graphs in FIG.~1.

\noindent  
{\mbox{ }\hfill\parbox{0.85\textwidth}{%
\epsfxsize=0.84\textwidth\epsfbox{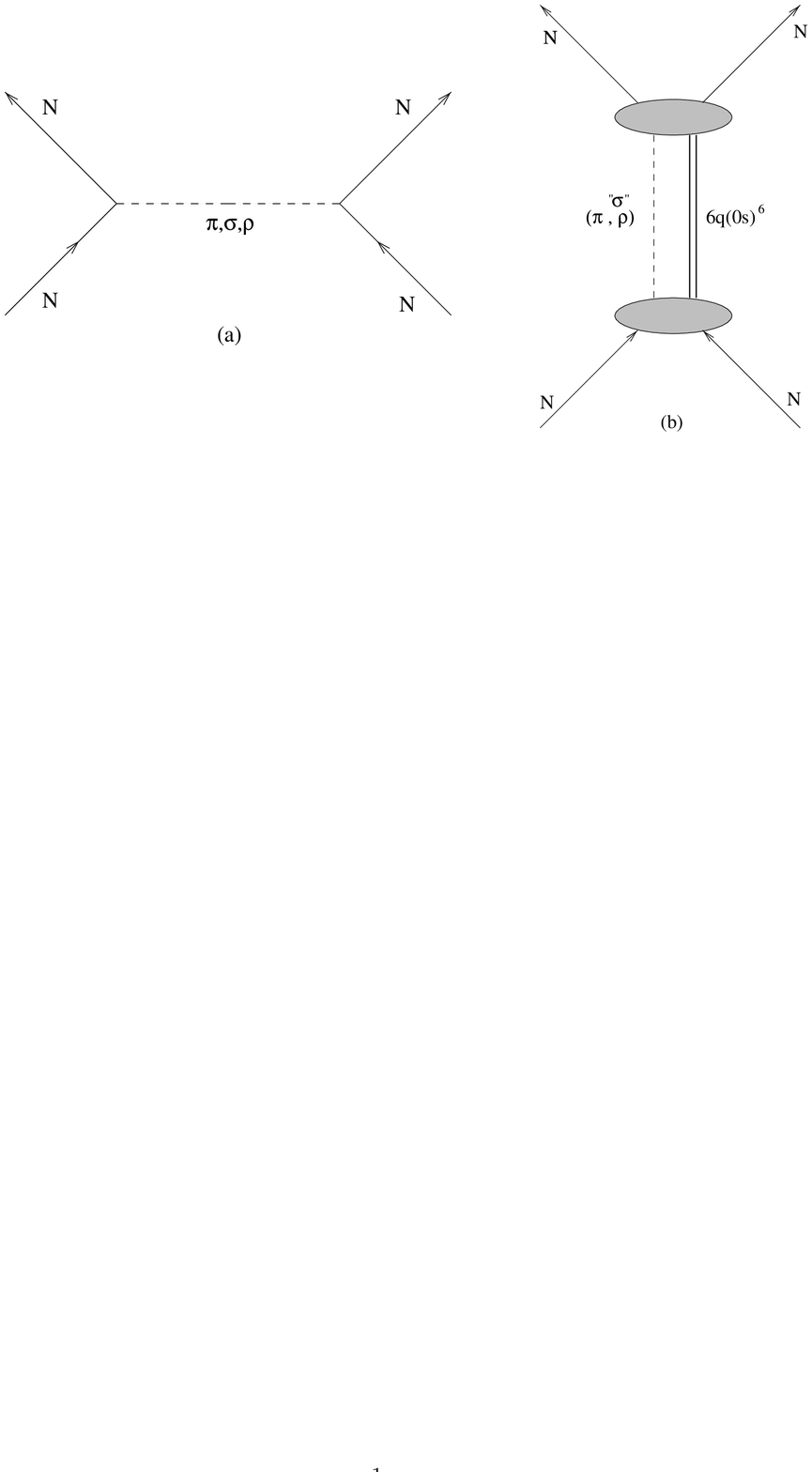}}}\\

{\bf FIG.1.} {\small The traditional t-channel meson-exchange 
mechanism (a) compared to the new $s$-channel "dressed" bag mechanism (b) for 
the $NN$ interaction.}\\ 
 
The Yukawa one-meson exchange mechanism displayed in FIG.~1(a)  is
confronted the new s-channel mechanism of the dressed bag 
intermediate  state in FIG.~1(b).  
The two pion state produced in the lower vertex in FIG.~1(b) 
is modified in the high density six-quark bag in which   chiral symmetry may
be at least partially restored. The "$\sigma$" or a similar 
"scalar-isoscalar meson" is assumed to exist only in a high density
environment and  not in the vacuum, contrary to the $\pi$ and $\rho$ mesons.
This mechanism can describe the short-range repulsion and   the medium range
attraction and replaces the $t$-channel exchange of   $\sigma$- and an
$\omega$-meson in conventional Yukawa-type $NN$ forces.  The short-range
repulsion arises here due to an additional requirement  for mutual
orthogonality of the $NN$- and 6q-channels (see Subsection 4.A).

Instead of the "$\sigma$"-meson in FIG.~1(b), other mesons like $\pi$ and  
$\rho$ can also be considered within this mechanism.  The contributions of
$\pi$, $\sigma$ and $\rho$ mesons will depend on the   total angular
momentum, spin-isospin and permutation symmetry of the   respective six-quark
state. Thus we adopt the $s$-channel quark-meson   intermediate states, the
transition amplitude being determined by $s$-channel   singularities in sharp
contrast to the Yukawa mechanism driven by   $t$-channel meson exchange (see
FIG.1(a)). Surely together with this specific   six-quark mechanism we take
into consideration also the traditional Yukawa   mechanism for $\pi$-,
$2\pi$- and $\rho$- (but not $\sigma$-) meson exchanges between   isolated
nucleons. However these meson-exchange contributions are essential   only at
the separations beyond the intermediate six-quark bag or in high   partial
waves ($L>3$). In the lowest partial waves, the intermediate dressed  
six-quark bag gives a dominating contribution for the total $NN$  
interaction. It is appropriate to refer henceforth to the present 
microscopic force model as a Moscow-T\"ubingen dressed bag model. 

\subsection{Short-range repulsion and orthogonality of the nucleon-nucleon 
and six-quark (bag) components} 
  
In our symmetry considerations we start from the well known results of  
previous works in this
field~\cite{Buch88,Obu79,Harv81,Oka83,Fae83,Kus91,Obu96,%
Kuk90,Kuk90Jap,Progr92,Yam86}. If one  assumes for the nucleon a wave
function of three constituent quarks with  a fully  symmetric spatial part
$[f_X]=[3]$ then the space (permutational)  symmetry of the six-quark system 
can be presented as follows: 
\begin{eqnarray*} 
\ [f_X]_{\rm even}=[6]+[42], \qquad \mbox{ for even-parity partial waves} 
\\ 
\ [f_X]_{\rm odd}=[51]+[33], \qquad \mbox{ for odd-parity partial waves}  
\end{eqnarray*} 
Further we adopt the nucleon wave function in the form 
\begin{equation} 
N(123)=|(0s)^3[3]_X[21]_{CS}S,T=1/2,\,\,1/2\rangle , 
\label{n123} 
\end{equation} 
where the coordinate part is the translationally-invariant harmonic 
oscillator (h.o.) state  
\begin{eqnarray} 
&|(0s)^3[3]_X\rangle =|0s(\rho_1)\rangle |0s(\rho_2)\rangle ,\qquad  
|0s(\rho_1)\rangle \sim e^{-\rho_1^2/4b^2},|0s(\rho_2)\rangle 
\sim e^{-\rho_2^2/3b^2},&\nonumber\\ 
&{\b\rho}_1={\b r}_1-{\b r}_2, 
\quad{\b\rho}_2=\frac{1}{2}({\b r}_1+{\b r}_2)-{\b r}_3&
\label{ho}
\end{eqnarray}  
and \{${\bf r}_1,{\bf r}_2,{\bf r}_3$\} are quark coordinates.
Then the translationally-invariant shell-model (TISM) configuration for  
six-quark states $\Psi_{6q}$ in the $NN$ overlap region can be written  
as follows (with restriction to   
configurations with only minimal numbers of h.o. quanta): 
\begin{eqnarray} 
\Psi_{6q} \to |(0s)^6[6]_X,[f_{CS}],L=0;ST\rangle  
+\sum_{f^{\prime}}C_{f^{\prime}}|(0s)^4(1p)^2[42]_X,[f^{\prime}_{CS}], 
L=0(2);ST\rangle ,\nonumber \\  
\mbox{for even waves 
(with $[f_{CS}]$=$[2^3]$ for ST=10 and $[2^21^2]$ for $ST$=01)} 
\label{even}
\end{eqnarray} 
and 
\begin{eqnarray} 
\Psi_{6q} \to |(0s)^5(1p)[51]_X,[f_{CS}],L=1;ST\rangle  
+\sum_{f^{\prime}}C_{f^{\prime}}|(0s)^3(1p)^3[3^2]_X,[f^{\prime}_{CS}], 
L=1(3);ST\rangle , \nonumber\\  
\mbox{for odd waves 
(with $[f_{CS}]$=$[2^21^2]$ for $ST$=00 and [321] for $ST$=11),} 
\label{odd}
\end{eqnarray} 
where $[f^{\prime}_{CS}]=[42],\,\,[321],\,\,[2^3],\,\,[31^3],\,\,[21^4]$ are 
all possible color-spin (CS) Young schemes for the inner product  
$[2^3]_C\circ[42]_S$ for $S$=1 and $[f^{\prime}_{CS}]=[2^3]_C\circ[3^2]_S= 
[3^2],\,\,[41^2],\,\,[2^21^2],\,\,[1^6]$ for $S$=0. 
 
Let us consider first the triplet $S$-wave $NN$ scattering, e.g. in the channel  
$L=0, ST=10$. In this case both allowed six-quark configurations  
\begin{eqnarray}  d_0=|(0s)^6[6]_X,[2^3]_{CS},L=0;ST=10\rangle  \mbox{ \
and}\nonumber\\ 
d^{L=0}_{f^{\prime}}=|(0s)^4(1p)^2[42]_X,[f^{\prime}_{CS}],L=0;ST=10
\rangle \,, 
\label{df}  
\end{eqnarray}  
correspond to state vectors of very
different nature: while the unexcited six-quark states $d_0$  include 
the states
with a maximal overlap all six quarks, the states with mixed symmetry  
$d^L_{f^{\prime}}$, $L$=0,  are those with two excited  $p$-shell quarks 
projected onto the $NN$ channel (with unexcited   nucleons) correspond to
cluster-like {\em nodal} $NN$ relative-motion   wave functions
$|\,2s(r)\rangle $(see e.g. \cite{Kus91}):  
\begin{eqnarray} 
\langle N(123)N(456)|d_0\rangle \,=\Gamma_{d_0}U^{NN}_{f_{0}}|\,0s(r)\rangle \,,  \nonumber\\ 
\langle N(123)N(456)|d_{f^{\prime}}^{L=0}\rangle \,= 
\Gamma_{d_{f^{\prime}}}U^{NN}_{f^{\prime}}|\,2s(r)\rangle ,
\label{prj} 
\end{eqnarray}   
where ${\b r}=\frac{1}{3}({\b r}_1+{\b r}_2+{\b r}_3)- 
\frac{1}{3}({\b r}_4+{\b r}_5+{\b r}_6)$ is the relative distance between 
two nucleons. In the $L=2$ case the projection onto the $NN$ channel  
\begin{equation}  \langle N(123)N(456)|d_{f^{\prime}}^{L=2}\rangle \,= 
\Gamma_{d_{f^{\prime}}}U^{NN}_{f^{\prime}}|\,2d(r)\rangle  
\label{prj2} 
\end{equation}  
leads to the cluster-like $2d$ $NN$ state. (We denote by the symbols
$f_0,\,f^{\prime}$ the following Young schemes:  
$f_0=\{[1^6]_{CST}\,,\,[2^3]_{CS}\}$ and 
$f^{\prime}=\{[f^{\prime}_{CST}]\,,\,[f^{\prime}_{CS}]\}$).  
In the
Eqs.~(\ref{prj}) and (\ref{prj2}) $U^{NN}_{f_0}$ and  $U^{NN}_{f^{\prime}}$ 
are overlaps   in the CST space  (see Table V in Appendix),
while  $\Gamma_{d_0}\equiv\Gamma(s^6[6]_X)=1$ and  
$\Gamma_{d_{f^{\prime}}^{L}}\equiv\Gamma(s^4p^2[42]_XL)= 
-\sqrt{\frac{4}{45}}$  ($L$=0,2) are trivial fraction parentage coefficients  (f.p.c.)
of the TISM.  

It is well known that both types of configurations are mixed if one  
assumes the basic $qq$ interactions to be only the OGE (or other effective  
color-dependent interactions)~\cite{Harv81,Oka83,Fae83,Kus91,Obu96,Myhr88}.  
(The states  $d^{L=0}_{f^{\prime}}$ and $d^{L=2}_{f^{\prime}}$ are also mixed
by effective $qq$ interactions and that gives rise to a new type of tensor 
force of quark origin, see Subsection 4.C.4). However, in a more microscopic 
treatment,  the
structure   of fully symmetric states $d_0$ ($ST$=10 or 01) should be
distinguished strongly from the mixed symmetry states
$d^{L=0}_{f^{\prime}}$ due to the enhanced quark density in non-excited
symmetric states. 

In fact, we are aware from  previous studies (see 
e.g.~\cite{Kun94,Br96,Rehb99}) based on   chiral  restoration effects in
multiquark systems either in strong color-electric fields  or in high density
nuclear matter within the framework of the Nambu-Jona-Lasinio model, that 
some phase transition may happen in increasing the quark density or  the
temperature of the system. This phase transition leads to a restoration   of
the  broken  chiral symmetry. Thus it is very plausible to assume that in
fully symmetric  $6q$-states there occurs a transition to (at least) a 
partial
restoration of  chiral symmetry which leads, in turn, to some decrease in
the mass difference of pions and scalar mesons (i.e. to some decrease of 
both the mass $m_{\sigma}$ and decay width $\Gamma_{\sigma\to 2\pi}$ of
the $\sigma$ meson~\cite{Kun94,Br96,Rehb99}). While  the mutual overlap of 
six quarks
in the mixed-symmetry states like  $|s^4p^2[42]_XLST\rangle$ should be much
smaller due to cluster-like structure of  these states (see below). Hence  the
structure of the multiquark states with maximal spatial symmetry should be
noticeably  different from states of mixed symmetry and the mixing of both
types of  states should be damped. 

The very important role which the $s^6$ bag surrounded by chiral fields
plays   for the  intermediate-range attraction in $NN$ interaction is
supported also from a very  different point of view.  In fact, very recently
it was established within chiral perturbation  
theory~\cite{Kai97,Kai98,Oset99} that in the $2\pi$-exchange diagrams
conventionally   associated to  "$\sigma$-meson" exchange\cite{Mach89} the
$\pi\pi$ $s$-wave correlation plays a  minor role while the excitation of
intermediate $\Delta$-isobars, in particular, the $\Delta\Delta$ channel 
in $S$-wave $NN$ interaction should be of prime  importance in the $NN$
interaction. But the $\Delta\Delta$ channel 
has a very high threshold ($\sim  600$~MeV in c.m. system)
and in low-energy $NN$ scattering (say, until $E_{\rm  lab}\simeq 300$~MeV)
this channel is strongly suppressed and thus the respective  intermediate
$\Delta\Delta$ state should be well localized at rather short  ranges
$r_{\Delta\Delta} \sim 0.5 \div 0.8 $~fm, i.e. well inside the overlap 
region of the two nucleons. 

	From  the other side, the six-quark wavefunction
$|(0s)^6[6]_X,L=0\rangle$,   being expanded into $N^*N^*$-components via
fractional parentage coefficients (f.p.c.)~\cite{Kus91,Obu96,Clust83}, has a
very significant $\Delta\Delta$-component. Thus, in the language of  the
quark model the fully symmetric six-quark configuration  
$|(0s)^6[6]_X,L=0\rangle$ (surrounded by $\pi$ + $\sigma + \rho$ fields)may 
replace very naturally the $\Delta\Delta$ intermediate-state channel in 
the traditional picture of $NN$ attraction. We should also stress here that
in the  above $\Delta\Delta$ channel the $\Delta$-isobars -- due to their
strong mutual overlap  --
should be  considered rather as "structural" deltas and thus this 
conclusion is in a very good agreement with our suggestion to replace the 
strongly closed   
$\Delta\Delta$ channel with our "dressed" $6q$ bag in a configuration  
$|(0s)^6[6]_X\rangle  + |(\pi\pi)\rangle$ (see FIG.~1(b) and below).   

Here we stress once again that we consider not the $t$-channel
meson-exchange  diagrams as in the Yukawa approach (FIG.~1(a)) but the
$s$-channel eigen-energy diagrams  where the energy and momenta of
intermediate $\pi$-, $\sigma$- and $\rho$-mesons  are taken into
consideration in {\em an explicit form} (see FIG.~1(b)). 

Another basic fact for our approach is the microscopic
calculations  for the $NN$ interaction based on a model six-quark 
Hamiltonian, especially the  results~\cite{Buch88,Oka83,Kus91,Yam86,Myhr88}, 
with the important conclusion: 

-- the coherent superposition of the {\it mixed symmetry} states  
$\sum_{f^{\prime}}C_{f^{\prime}}d_{f^{\prime}}$ obtained from the 
$6q$-Hamiltonian being projected onto the $NN$ channel yields a
large  reduced width for the channel with two unexcited nucleons and a  small
reduced width for the channel with two $\Delta$'s, while the fully   symmetric
eigenstate $d_0$ is  projected onto both $NN$ and $\Delta\Delta$ channels
approximately with the  same reduced width (or probability). This  result
makes it possible to identify the above {\em mixed symmetry} $6q$ states as  
related mainly to the proper $NN$ channel whereas the {\em fully 
symmetric}  states
$d_0$ in both channels $ST$=01 and 10 (see Eq.~(\ref{df})), which are  
orthogonal to the former states, represent non-nucleonic bag-like  
states~\cite{Kus91,Kuk90,Kuk90Jap,Progr92}%
\footnote{This statement in the form of very plausible conjecture
has been suggested first as early as in 1986 \cite{Obu86,LIYAF86}}. 

Thus the total six-quark wave function can be divided according to the 
above consideration into two mutually orthogonal components: 
cluster-like configurations with a node and bag-like configurations
without a node: 
\begin{equation} 
\Psi_{6q}=\Psi_{\rm bag} +\Psi_{\rm cluster}, 
\label{pr1} 
\end{equation} 
\begin{equation} 
\langle \Psi_{\rm bag} |\Psi_{\rm cluster}\rangle =0. 
\label{pr2} 
\end{equation} 

This separation of $\Psi_{6q}$ into two mutually orthogonal components have
been  employed extensively  in developing the so-called Moscow $NN$ 
potential~\cite{Kuk90,Kuk90Jap,Progr92,ru73} which is rather
successful in describing the $NN$ interaction on the  level of a
semi-phenomenological potential model~\cite{PRC99,Kuk98,Moscow,KuFaes98,%
LIYAF86} and offered many 
interesting predictions and explanations for some long-standing puzzles. In 
particular, the orthogonality condition~(\ref{pr2}) leads in low partial
waves  to almost stationary nodes in $NN$ scattering wave functions 
\cite{Moscow,KuFaes98,LIYAF86} exactly
at the  place of the repulsive hard core in old phenomenological $NN$
potentials, like  RHC (Reid Hard Core) etc. And thus this natural orthogonality condition
plays the role of the repulsive core (or of its main part) at low partial waves
in our approach. Due to the effect of orthogonality the strength of the
repulsive core caused by $\omega$-meson exchange can be noticeably reduced 
\cite{KuFaes98}
and the $\omega NN$ coupling constant can be taken around the values 
$g^2_{\omega NN}/4\pi \simeq 5$ in nice agreement with $SU(3)$ predictions%
\footnote{The authors are much obliged to
Prof.~Gerry~Brown who attracted their attention just to this  aspect of the
problem related to the $\omega NN$ coupling constant  puzzle.}.

\subsection{"Dressed" six-quark bags at short and intermediate $NN$ ranges} 
 
In the present work we make a next step in this direction and develop a 
microscopic model which can replace the intermediate and short range part of 
existing OBE potentials and which can simultaneously explain the  
enhanced meson-exchange vertices (large cutoff parameters) at short ranges 
in the traditional OBE approach to the $NN$ interaction. 

The model uses an analogy between an excited atom or molecule and a
multiquark bag  with excited $p$-shell quarks like $s^4p^2$. Similarly to
e.m. transitions in atomic  physics we assume that each $p$-shell quark emits
a pion -- a "quantum"   of the chiral field -- in the process of its
transition from the $p$- to $s$-shell (see FIG.~2).  

\epsfxsize=0.5\textwidth 
\centerline{\epsfbox{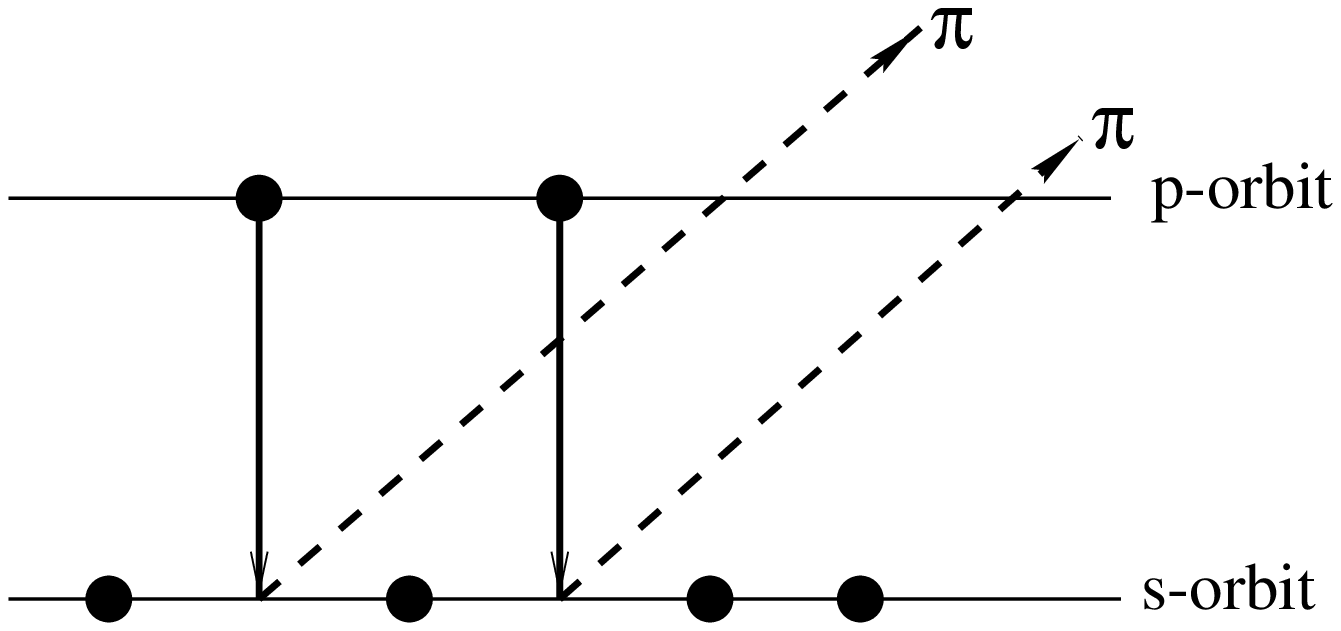}} 
\nopagebreak
{\bf FIG.2.} {\small Schematic representation of the 
two-pion emission in the transition of two $p$-shell quarks to an $s$
orbit.}\\[0.2cm]

The transition $NN^{L=0,2}(s^4p^2) \to 6q(s^6) +\sigma$ involving
deexcitation of two quarks from the $p$-orbits in configurations $s^4p^2$
(and also the   similar  transition for $P$-waves $NN^{L=1}(s^3p^3) \to
6q(s^5p) +\sigma$) can be  described as subsequent emission of two pions,
each of them deexciting one  $p$-orbit, followed by the formation of a
$\pi\pi$ resonance inside the $6q$ bag giving  the $\sigma$- or $\rho$-mesons.     
The mechanism of elastic $S$-wave $NN$ scattering via an
intermediate symmetric  $6q$ state $|(0s)^6[6]_X,L=0\rangle +|(2\pi)\rangle$
can be displayed by the graph   in  FIG.~3, where we assume the emission of
intermediate $\sigma$- and $\rho$-mesons  and their subsequent absorption 
which take place on a diquark (correlated quark pair) inside the  $6q$ bag. 

\epsfxsize=0.7\textwidth 
\centerline{\epsfbox{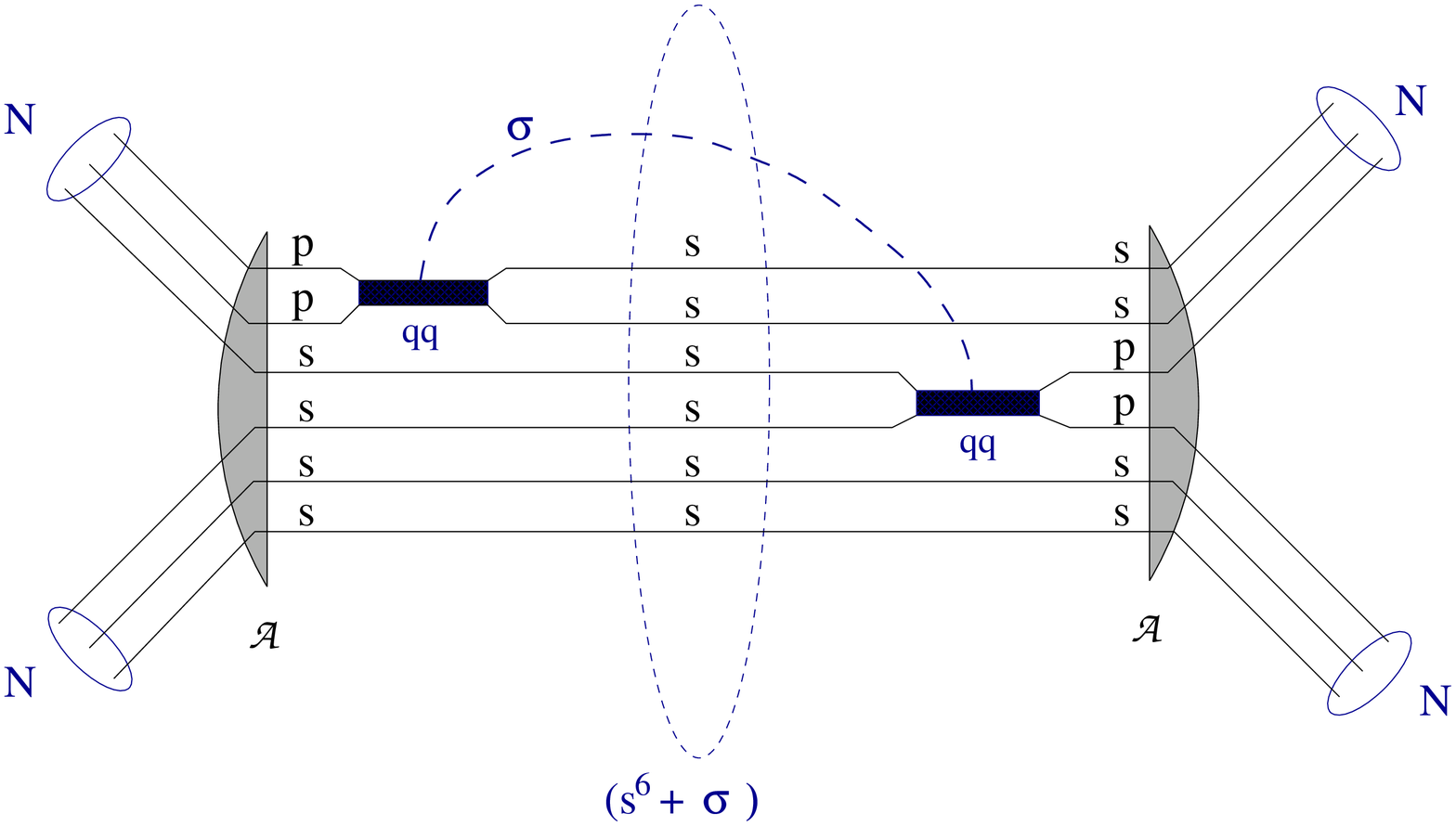}} 
{\bf FIG.3} {\small The graph illustrating the $\sigma$- (or 
$\rho$-) meson emission and subsequent absorption by diquark pairs in  
the intermediate six-quark bag-like state.}\\[0.2cm]
 
Another variant of this process is displayed in FIG.~4 where the $\sigma$- and 
$\rho$-mesons are formed from two independent pions in the process of their 
interaction inside the bag.

\epsfxsize=0.7\textwidth 
\centerline{\epsfbox{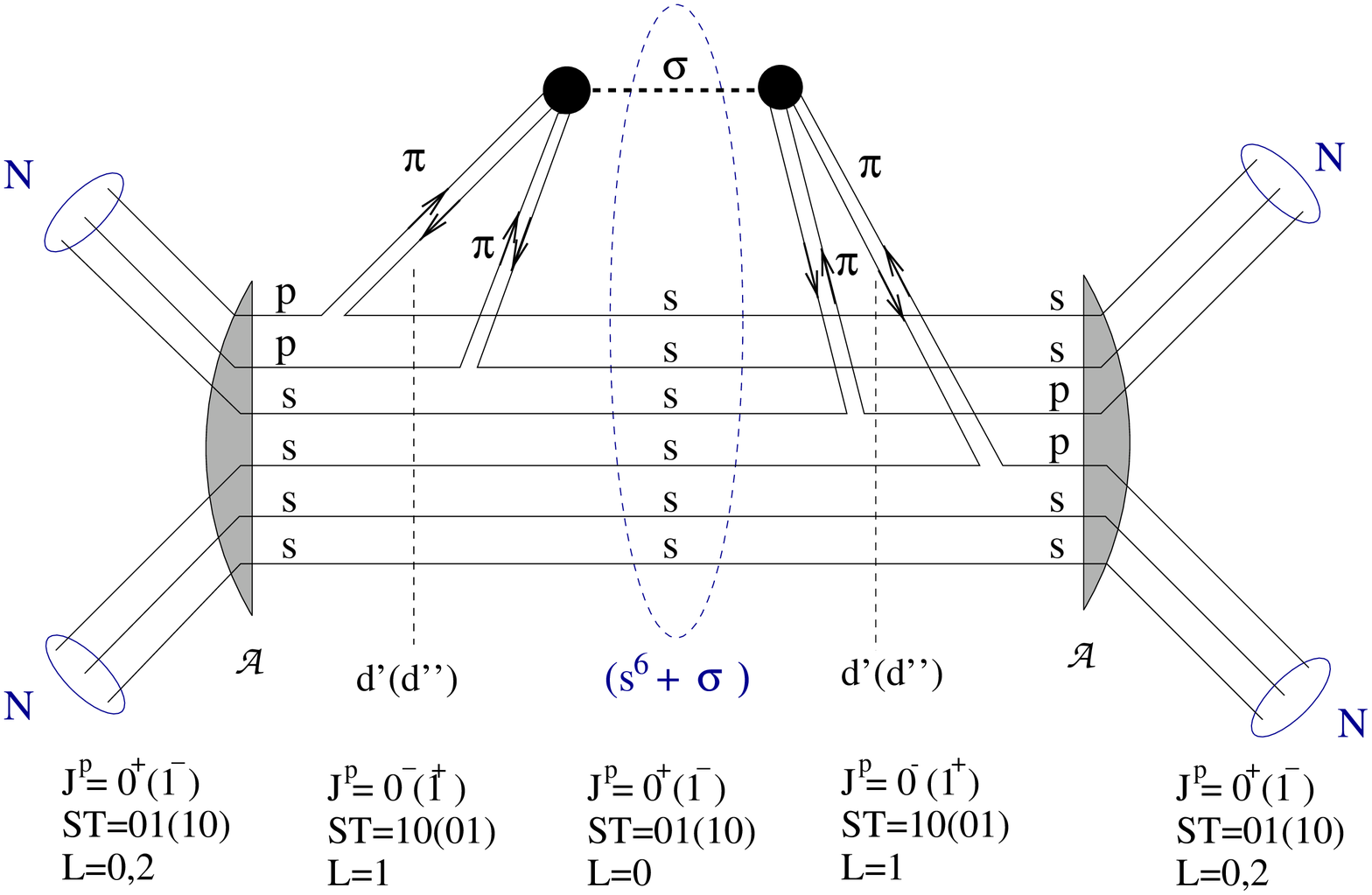}} 
{\bf FIG.4.} {\small The graph illustrates two sequential 
$\pi$-meson emissions and absorptions via an intermediate $\sigma$- (or 
$\rho$-) meson and the generation of a six-quark bag.} \\[0.2cm]

In the graph (FIG.~4) the pions are created in $s$-waves due to 
conservation of parity and angular momentum. The intermediate six-quark 
configuration $s^5p[51]_X$ (denoted by vertical dashed line in the graph) 
for even  waves $L=0,2$ in the $NN$ channel has fixed quantum numbers  
which are determined by those in the initial ($NN$) and final 
("dressed" bag $6q+\sigma$) states.  
The intermediate (after the first pion emission) state in the channel 
$ST=01,\,\,J^P=0^+$ has quantum numbers of 
the so-called $d^{\prime}$-dibaryon (see, e.g.~\cite{Obu99,Wa95,Bu98}):
\begin{eqnarray}
|d^{\prime}\rangle =|(0s)^5(1p)[51]_X[321]_{CS},LST=110, J^P=0^-\rangle.
\label{d1}
\end{eqnarray}
The transition into the channel $ST=10,\,\,J^P=1^-$ goes via an intermediate 
state $d^{\prime\prime}$, 
which is a partner of the $d^{\prime}$ with $S\to T$ interchanged: 
\begin{eqnarray}
|d^{\prime\prime}\rangle =|(0s)^5(1p)[51]_X[321]_{CT},LST=101, 
J^P=1^+\rangle.\nonumber\\ 
=|(0s)^5(1p)[51]_X[2^21^2]_{CS},LST=101, J^P=1^+\rangle. 
\label{d2}
\end{eqnarray}
 
It should be noted that both states $d^{\prime}$ and $d^{\prime\prime}$  
have no direct coupling with the $NN$ channel due to the requirement 
of antisymmetrization and due to this feature they have been considered 
in some previous works 
as candidates for narrow dibaryon resonances (see e.g. the discussion on  
$d^{\prime}$ in Refs.~\cite{Obu99,Wa95,Bu98,Obu97}). Both transitions   
$NN^{L=0}(ST\!=\!01,\!J^P\!=\!0^+)\to d^{\prime} +\pi$  
and $NN^{L=0,2}(ST\!=\!10,J^P\!=\!1^-) \to d^{\prime\prime}+\pi$ proceed with  
a spin and isospin flip of the quark emitting the S-wave $\pi$-meson. 
The further decays $d^{\prime}\to d_{0}(ST\!=\!01)+\pi$ and  
$d^{\prime\prime}\to d_{0}(ST\!=\!10)+\pi$ proceed with the spin-isospin-flip 
mechanism as well (for details see Refs.~\cite{Obu99,Obu97}). 
However in case of the coupling mechanism shown in FIG.~4, the existence 
of such narrow dibaryon resonances in the transition $s^4p^2\to s^6+2\pi$ 
can destroy the coherence between the emissions of two pions and thus should 
be incompatible with such a mechanism.
 
\subsection{Quark-model calculation of the transition operator from the $NN$ 
cluster channel to the "dressed"-bag state} 

From a general point of view the calculation of the driving term 
corresponding to the transition displayed in FIG.~3 which includes a direct 
coupling of $\sigma$- and $\rho$-mesons with diquarks in the multiquark bag
looks much more preferable and straightforward than the two-step mechanism of
subsequent $\pi$-emissions and absorptions shown in FIG.~4. Unfortunately now
we have no reliable estimates neither for the dynamics and probabilities of
diquarks in the multiquark bag nor for the strength of coupling
$(qq)\to\sigma+(qq)$ (or $(qq)\to\rho+(qq)$). Therefore we postpone this
estimation for the future work and present here the calculation only for the 
two-step mechanism of the coupling of a $\sigma$ to a six-quark bag.

The total amplitude for the transition 
$NN^{L=0,2}(s^4p^2) \to 6q(s^6) +\sigma$ to the "dressed" bag  
is calculated as a contribution of the triangle graph shown in FIG.~5. Each 
of the two lower vertices are calculated in the framework of the  
phenomenologically successful $^3P_0$ quark-pair-creation  
model (QPCM)~\cite{Micu69} which appears to be very  
useful~\cite{Godf85,Kok87,Cap93} in the flux-tube picture of  
hadrons~\cite{Isgur85}.

\epsfxsize=0.5\textwidth 
\centerline{\epsfbox{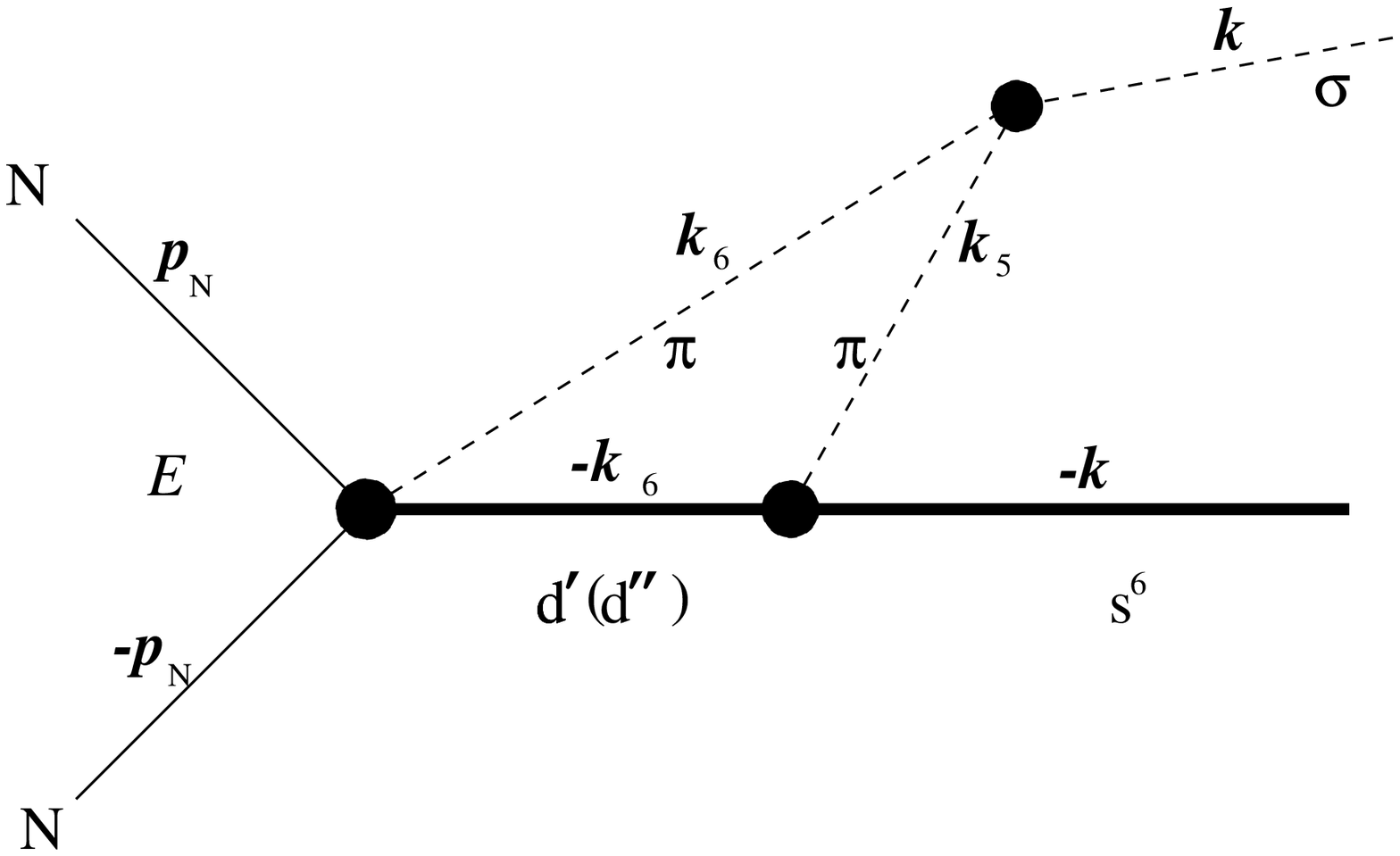}} 
{\bf FIG.5.} {\small Kinematical variables in the 
triangle diagram corresponding to the $\sigma$- (or $\rho$-) meson 
creation by two $\pi$-mesons formed in the transition of two 
$p$-shell quarks to the $s$-orbit (see also FIG.~4).}\\[0.2cm]

\subsubsection{Elementary vertex operators} 
 
In the QPCM the transition operator for the emission of the pion 
$\pi^{\lambda}$ 
($\lambda=0,\pm$) by the sixth quark in a six-quark system can be written as 
(see Ref.~\cite{Obu99} for details) 
\begin{eqnarray} 
H^{(6)}_{\pi qq}({\b k}_6;{\b\rho}_5,{\b\rho}^{\prime}_5)= 
v\,\tau_{-\lambda}^{(6)}e^{i{5 \over 6}{\bf k}\cdot{\b\rho}^{\prime}_5} 
\hat O^{(6)}({\b\rho}_5,{\b\rho}^{\prime}_5) 
\,{\b\sigma}^{(6)}\!\cdot\!\left[{{\omega_{\pi}} \over {2m_q}} 
\left({2 \over i}{\b\nabla}_{\rho_5}+{5 \over 6}{\bf k}_6\right)+ 
\left(1+{{\omega_{\pi}} \over {12m_q}}\right){\bf k}_6\right], 
\label{H} 
\end{eqnarray} 
where the non-local factor is proportional to the pion wave function 
 
\begin{eqnarray} 
\hat O^{(6)}({\b\rho}_5,{\b\rho}^{\prime}_5)= 
e^{-i{1 \over 2}{\bf k}_6\cdot({\b\rho}_5-{\b\rho}^{\prime}_5)} 
\Psi_{\pi}({\b\rho}_5\!-\!{\b\rho}^{\prime}_5). 
\label{O} 
\end{eqnarray} 
Here ${\b\rho}_5$ and ${\b\rho}_5^{\prime}$ are the relative coordinates of  
the 6th quark in the initial and final states respectively, viz.  
$${\b \rho_5}=\frac{1}{5}({\b r}_1+{\b r}_2+\dots+{\b r}_5)-{\b r}_6\,, 
\qquad {\b \rho^{\prime}_5}=\frac{1}{5}({\b r}_1+{\b r}_2+\dots+{\b r}_5) 
-{\b r^{\prime}}_6\,.$$  
We use shell-model quark configurations for the pion and the $\sigma$ meson
\begin{eqnarray} 
\pi^{\lambda}=\vert s\bar s[2]_XLST=001\,T_z\!=\!\lambda\,\, 
J^P\!\!=0^-\rangle \nonumber\\  
\sigma=|s^2\bar{s}^2[4]_X,LST\!=\!000,J^P=0^+\rangle , 
\label{qconf} 
\end{eqnarray} 
that imply Gaussian wave functions:   
\begin{eqnarray} 
&\Psi_{\pi}({\b \rho_{\pi}})\sim \exp(-\rho_{\pi}^{2}/4b_{\pi}^2),&
\nonumber\\ 
&\Psi_{\sigma}({\b \rho}_{\sigma};{\b \rho}_{\pi}(12),\,{\b \rho}_{\pi}(34)) 
\sim \exp\left[-\rho_{\sigma}^{2}/2b_{\sigma}^2-
\rho_{\pi}^{2}(12)/4b_{\pi}^2- 
\rho_{\pi}^{2}(34)/4b_{\pi}^2\right],&\nonumber\\ 
&\mbox{where}\quad{\b \rho}_{\sigma}=\frac{1}{2}({\b r}_1+{\b r}_2)-  
\frac{1}{2}({\b r}_3+{\b r}_4) 
\quad \mbox{and}\quad {\b \rho}_{\pi}(ij)={\b r}_i-{\b r}_j&
\label{pisi} 
\end{eqnarray} 
 
In the limit of a point-like pion ($b_{\pi}\to 0$) the 
operator (\ref{H}) goes to the standard pseudo-vector (PV) quark-pion  
coupling and the phenomenological constant $v$ in Eq.~(\ref{H}) becomes 
the PV coupling constant:
\begin{eqnarray} 
v=-i\,{{f_{\pi qq}} \over {m_{\pi}}} 
\frac{1}{(2\pi)^{3/2}(2\omega_{\pi})^{1/2}}, 
\label{v} 
\end{eqnarray} 
where $f_{\pi qq}$ should be normalized to the well known pion-nucleon 
PV coupling constant $f_{\pi qq}=\frac{3}{5}f_{\pi NN}$. 
 
The $\pi +\pi \to \sigma$ transition amplitude is determined 
(see details in Ref.~\cite{Guts94})  
to be proportional to the overlap  
of the two wave functions of the $\pi$-meson with the $\sigma$-meson 
wave function: 
\begin{eqnarray} 
\langle \pi(12)\pi(34)|H_{\pi\pi\sigma}({\b k},{\b k^{\prime}})|\,\sigma\rangle  
=g_{\pi\pi\sigma}F_{\pi\pi\sigma}(({\b k}-{\b k^{\prime}})^2), 
\quad F({\b q}^2)=\exp(-\frac{1}{2}q^2b_{\sigma}^2), 
\label{sigm} 
\end{eqnarray}              
where the transition operator $H_{\pi\pi\sigma}$ contains 
a phenomenological constant G:  
$$H_{\pi\pi\sigma}({\b k},{\b k^{\prime}})=  
G\exp[i(\frac{{\b k}-{\b k^{\prime}}}{2})\cdot{\b\rho}_{\sigma}],$$  
and the effective coupling constant $g_{\pi\pi\sigma}$ in Eq. (\ref{sigm})  
is proportional to the f.p.c. $\Gamma_{\pi\pi\sigma}$ for decomposing the  
two pion $q\bar q$ states in the configuration of Eq.~(\ref{qconf}):
$g_{\pi\pi\sigma}=G\Gamma_{\pi\pi\sigma}$. 
The coefficient $\Gamma_{\pi\pi\sigma}$ includes contributions from both  
CST (color, spin, isospin) and coordinate overlaps. 
 
\subsubsection{Transition operator: Decomposition in basis of six-quark  
configurations} 
 
The total transition amplitude $NN^{L=0,2}(s^4p^2) \to d'(d'')+\pi  
\to 6q(s^6)+\sigma$ can be written as 
\begin{eqnarray} 
A^{L=0(2)}_{NN\to d_0+\sigma}(E;{\b k})=\int d^3r\,\Psi_E^{L=0(2)}({\b r})\,
\Omega_{NN\to d_0+\sigma}(E;{\b r},{\b k})\,, 
\label{A} 
\end{eqnarray} 
where $\Psi_E^{L=0(2)}({\b r})$ is the cluster part of the $NN$ wave function 
in the sense of Eqs.~(\ref{pr1}) - (\ref{pr2}) and 
$E=2m_N+\frac{p_N^2}{m_N}$. 
The transition operator $\Omega_{NN\to d_0+\sigma}(E;{\b r},{\b k})$ 
in Eq.~(\ref{A}) is the contribution of the triangle diagram of FIG.~5. 
Note that the standard momentum representation of the amplitude 
could be obtained by inserting the plane-wave decomposition of the unit operator 
$I=\int d^3p_N|{\b p}_N\rangle \langle {\b p}_N|$ into Eq.~(\ref{A}):
\begin{eqnarray} 
T_{NN\to d_0+\sigma}(E;{\b p}_N,{\b k})=\int\frac{e^{i{\b p}_N 
\cdot {\b r}}}{\sqrt{(2\pi)^3}}\,\,\Omega_{NN\to d_0+\sigma}(E;{\b r},{\b k})
\,d^3r
\label{t} 
\end{eqnarray} 
 
We start from the quark-meson diagram of FIG.~4 and project the six-quark 
state in the left part of 
the diagram onto the two-nucleon clusters of the initial state. The full 
expression for the transition operator $\Omega_{NN\to d_0+\sigma}$ 
can be written as an integral of 
the elementary six-quark  
transition amplitudes over both inner coordinates of quark clusters (viz. 
N(123), N(456), and also $\pi$ and $\sigma$) and the pion momenta  
${\b k}_5=\frac{{\b k}+{\b q}}{2}$ and   
${\b k}_6=\frac{{\b k}-{\b q}}{2}$ in the triangle diagram:  
\begin{eqnarray} 
&\Omega_{NN\to d_0+\sigma}(E;{\b r},{\b k})= 
15 { \displaystyle\int d^3k_5\,\int d^3k_6\,
\delta({\b k}_5+{\b k}_6-{\b k})}&\nonumber\\ 
&{\times}\sqrt{10}\,\langle N(123)N(456)|H_{\pi qq}^{(6)}({\b k}_6) 
|\,d^{\prime}(d^{\prime\prime})\rangle\,
G^{(0)}_{\pi d^{\prime}(d^{\prime\prime})}(E;{\b k}_6)\,
\langle d^{\prime}(d^{\prime\prime})|H_{\pi qq}^{(5)}({\b k}_5)|\,d_0\rangle & 
\nonumber\\ 
&{\times} G^{(0)}_{2\pi d_0}(E;{\b k}_5,{\b k}_6,)\,
\langle 2\pi|H_{\pi\pi\sigma}({\b k}_5,{\b k}_6)|\,\sigma\rangle & 
\label{om} 
\end{eqnarray} 
where $H^{(6)}_{\pi qq}$ ($H^{(5)}_{\pi qq}$) is the vertex operator  
(\ref{H}) of the effective quark-pion coupling for the 6-th (5-th) quark 
in the diagram of FIG.~4 and  $\sqrt{10}$ is a combinatorial  
factor ($\sqrt{\frac{6!}{3!3!2!}}$) in projection  
of the six-quark amplitude onto the $3q-3q$ cluster channel. 
In Eq.~(\ref{om}) $G^{(0)}_{\pi d^{\prime}}$ and  
$G^{(0)}_{2\pi d_0}$ are free Green functions for $\pi+d^{\prime}$ 
and $2\pi+d_0$ systems\footnote{Due to large masses of the intermediate
six-quark states $d_0$ and $d^{\prime}(d^{\prime\prime})$ and also the
$\sigma$- and $\rho$-mesons we use for them the nonrelativistic kinematics
while for intermediate $\pi$-mesons we use the relativistic kinematics}
\begin{eqnarray} 
&G^{(0)}_{\pi d^{\prime}}(E;{\b k}_6)= 
\left [E-m_{d^{\prime}}-\frac{k_6^2}{2m_{d^{\prime}}} 
-\omega_{\pi}(k_6)\right]^{-1},\qquad 
\omega_{\pi}(k_6)=\sqrt{m_{\pi}+k_6^2}\,,&\nonumber\\ 
&G^{(0)}_{2\pi d_0}(E;{\b k}_5,{\b k}_6,)= 
\left[E-m_{d_0}-\frac{({\b k}_5+{\b k}_6)^2}{2m_{d_0}} 
-\omega_{\pi}(k_5)-\omega_{\pi}(k_6)\right]^{-1}& 
\label{G} 
\end{eqnarray} 
and the numerical factor in front of the integral in the r.h.s. of  
Eq.~(\ref{om}) is the number of $qq$ pair in the six-quark system. 
 
In calculation of the amplitude Eq.~(\ref{A}) it is reasonable first
to project  
the initial $NN$ state onto the basis of six-quark configurations by 
inserting the unit operator  
\begin{eqnarray} 
I=\sum_{n,f}\vert n,f\rangle \langle n,f\vert  
\label{nn} 
\end{eqnarray} 
into the first vertex matrix element in the integrand of Eq.~(\ref{om}).  
Here we employ the full shell-model basis of six-quark configurations with  
quantum numbers  
of the initial $NN$ state $LST,\,J^P$, e.g. 
\begin{eqnarray} 
\vert n,f\rangle =\vert s^{n_s}p^{n_p}[f_X]\,[f_{CS}]LST,\,J^P\rangle \,, 
\label{n} 
\end{eqnarray} 
where symbols $n$ and $f$ are defined as $n$=\{$n_s$,$n_p$\}, 
$f=\{[f_X],\,[f_{CS}]\}$. In case of emission of $s$-wave pions, the 
excited six-quark configurations $d^{L=0(2)}_{f^{\prime}}$ in the 
sum (\ref{nn}) are only important (while the bag-like configuration  
$d_0$ does not 
contribute to the amplitude (\ref{A}) because of the orthogonality 
condition (\ref{pr2}) for the wave function $\Psi_E({\b r})$). Thus one can  
write the following decomposition of the vertex matrix element  
$N+N\to d^{\prime}(d^{\prime\prime})$ in the integrand of Eq.~(\ref{om}): 
\begin{eqnarray} 
&\sqrt{10}\langle N(123)N(456)|H^{(6)}_{\pi qq}({\b k}_6) 
|d^{\prime}(d^{\prime\prime})\rangle &\nonumber\\ 
&=\sum\limits_f\sqrt{10}\langle N(123)N(456)|d_f^{L=0(2)}\rangle \, 
\langle d_f^{L=0(2)}|H^{(6)}_{\pi qq}({\b k}_6)|d^{\prime}
(d^{\prime\prime})\rangle & 
\label{sprj} 
\end{eqnarray} 
and use the overlap factors from Eqs.~(\ref{prj}) and (\ref{prj2}). 
As a result, the
shell-model matrix elements of the vertex operator (\ref{H}) will only  
contribute to the transition amplitudes of Eqs.~(\ref{om}) and (\ref{A}).   
 
All the matrix elements of interest are calculated based on the 
f.p.c. technique \cite{Harv81,Kus91,Obu96,Obu99} 
(see the details  of calculation in Appendix) 
and are reduced to the standard form  
of the vertex matrix element as a product of a vertex constant  
$vf_{\pi AB}$, of a form factor $F_{\pi AB}(k_i)$ and of a kinematical 
factor  $\omega_{\pi}(k_i)/m_qb$ (see Ref.~\cite{Obu99} for details): 
\begin{eqnarray} 
\langle d_f^{L=0(2)}|H^{(6)}_{\pi qq}({\b k}_6)|d^{\prime}\rangle = 
v\frac{\omega_{\pi}(k_6)}{m_qb}f_{\pi d_fd^{\prime}}^{L} 
F^{L}_{\pi d_fd^{\prime}}(k_6^2)\,, 
\nonumber\\ 
\langle d^{\prime}|H^{(5)}_{\pi qq}({\b k}_5)|d_0\rangle = 
v\frac{\omega_{\pi}(k_5)}{m_qb}f_{\pi d_0d^{\prime}} 
F_{\pi d_0d^{\prime}}(k_5^2). 
\label{fF} 
\end{eqnarray} 
Form factors $F_{\pi d_fd^{\prime}}(k_6^2)$ and  
$F_{\pi d_0d^{\prime}}(k_5^2)$ in Eq.~(\ref{fF}) do not depend on the index 
$f$ of configurations $d_f$ and have in shell-model representation the form:   
\begin{eqnarray} 
F^{L}_{\pi d_fd^{\prime}}(k_6^2)=(1+a_L\frac{5k_6^2b^2}{24}) 
e^{-5k_6^2b^2/24}, \qquad  
F_{\pi d_0d^{\prime}}(k_5^2)=e^{-5k_5^2b^2/24} 
\label{FL} 
\end{eqnarray} 
where $a_{L=0}=\frac{1}{3}$ and $a_{L=2}=-\frac{2}{3}$.  
 
\subsubsection{Calculation of the transition operator} 
 
Substituting the vertex amplitudes (\ref{sigm}) and (\ref{fF}) into 
Eqs.~(\ref{om}) and (\ref{sprj}) one obtains the following simple expression  
for the transition operator (\ref{om}) in case of S and D partial waves in 
the initial $NN$-states: 
\begin{eqnarray}  
&\Omega_{NN\to d_0+\sigma}^{L=0}(E;{\b r},{\b k})=  
g_0e^{-5k^2b^2/48}D^L(E,{\b k})|2s(r)\rangle \,,&\nonumber\\ 
&\Omega_{NN\to d_0+\sigma}^{L=2}(E;{\b r},{\b k})=  
g_2e^{-5k^2b^2/48}D^L(E,{\b k})|2d({\b r})\rangle \,,&\nonumber\\
&\Omega_{NN\to d_0+\sigma}=\sum\limits_{L=0,\,2} 
\Omega_{NN\to d_0+\sigma}^{L}&
\label{om1} 
\end{eqnarray}  
where the values $g_L$ are effective strength constants of transitions from (cluster) 
$NN$ states to the "dressed" bag configuration $N+N\to d_0+\sigma$. The 
above calculation gives within the quark model the values $g_L$: 
\begin{eqnarray} 
g_{L}=15\frac{f_{\pi qq}^2}{m_{\pi}^2}\frac{1}{m_q^2b^2} 
f_{\pi d_0d^{\prime}}g_{\pi\pi\sigma}\sum\limits_f 
f_{\pi d_fd^{\prime}}^{L}\Gamma_{d_f}U^{NN}_f\,,\,\,\,\,L=0,\,2, 
\label{gL} 
\end{eqnarray}  
where the coefficients $\Gamma_{d_f}$ and $U^{NN}_f$ are defined according to  
Eqs.~(\ref{prj}) and (\ref{prj2}) while the vertex constants  
$f_{\pi d_0d^{\prime}}$ and $g_{\pi\pi\sigma}$ are taken from  
Eqs.~(\ref{sigm}) and (\ref{fF}). 
 
Function $D^L(E,{\b k})$ is obtained by an integration in Eq.~(\ref{om}) 
over inner momenta (the contribution of the triangular diagram  
of FIG.~5). By denoting the integration variable 
\begin{eqnarray} 
{\b q}={\b k}_5-{\b k}_6 
\label{kq} 
\end{eqnarray} 
one can rewrite the integral in the form 
\begin{eqnarray} 
&D^L(E,k)={ \displaystyle\int d^3q\omega^{1/2}_{\pi}
(\frac{{\b k}-{\b q}}{2}) 
\omega^{1/2}_{\pi}(\frac{{\b k}+{\b q}}{2}) 
[1+\frac{5}{24}a_{\scriptscriptstyle L}
(\frac{{\b k}-{\b q}}{2})^2b^2)]\,e^{-q^2B^2}}& 
\nonumber\\ 
&{\times} \left[m_{d_0}\!+\!\frac{k^2}{2m_{d_0}}\!+\! 
\omega_{\pi}(\frac{{\b k}-{\b q}}{2})+ 
\omega_{\pi}(\frac{{\b k}+{\b q}}{2})-E\right]^{-1} 
\left [m_{d^{\prime}}\!+\! 
\frac{1}{2m_{d^{\prime}}}(\frac{{\b k}-{\b q}}{2})^2\!+\! 
\omega_{\pi}(\frac{{\b k}-{\b q}}{2})-E\right]^{-1},& 
\label{D} 
\end{eqnarray}  
where ${\b k}={\b k}_5+{\b k}_6$ (the $\sigma$ meson momentum) is fixed. 
The result of integration depends on the absolute value of the 
vector ${\b k}$ but not on its direction, i.e. $D^L(E,{\b k})=D^L(E,k)$. 
Moreover, one can substitute useful equalities:
\begin{eqnarray} 
&\omega_{\pi}(({\b k}\!-\!{\b q})/2)+ 
\omega_{\pi}(({\b k}\!+\!{\b q})/2)= 
\omega_{\pi}((k\!-\!q)/2)+ 
\omega_{\pi}((k\!+\!q)/2)& 
\nonumber\\ 
&\omega_{\pi}(({\b k}\!-\!{\b q})/2) 
\omega_{\pi}(({\b k}\!+\!{\b q})/2)= 
\omega_{\pi}((k\!-\!q)/2) 
\omega_{\pi}((k\!+\!q)/2)& 
\label{som} 
\end{eqnarray} 
to reduce the angular dependence of the integrand.  
Note that because of the Gaussian factor $e^{-q^2B^2}$  
in the integrand with  
$$ 
B^2=5b^2/48+b_{\sigma}^2/2 
$$  
the integral in Eq.~(\ref{D}) is convergent. Hence, at small values of $kb$,  
one can use  the following decomposition for $D^L(E,k)$: 
\begin{eqnarray} 
D^L(E,k)=D_{0}(E,k)+\sum\limits_{n=1}^{}(k^2b^2)^{n} D^L_{2n}(E) 
\label{DD} 
\end{eqnarray} 
where $D_{0}(E,k)$ is obtained by neglecting $a_{\scriptscriptstyle L}$ 
in the numerator 
of the integrand. Then by making use a nonrelativistic approximation  
$\omega_{\pi}({\b p})\approx\omega_{\mbox{\small nrel}}=m_{\pi}+p^2/2m_{\pi}$ 
in the last factor of the integrand one can carry out 
the integration over angular variables in $D_0(E,k)$:            
\begin{eqnarray} 
D_{0}(E,k)=-\,\,\frac{8\pi m_{d^{\prime}\pi}}{k}\int\limits_0^{\infty} qdq 
\frac{\omega^{1/2}_{\pi}(\frac{k-q}{2}) 
\omega^{1/2}_{\pi}(\frac{k+q}{2})\,e^{-q^2B^2}} 
{m_{d_0}+\frac{k^2}{2m_{d_0}}+\omega_{\pi}(\frac{k-q}{2})+
\omega_{\pi}(\frac{k+q}{2})-E}\, 
\ln\left|\frac{1+\frac{(k-q)^2}{8m_{d^{\prime}\pi}\Delta E}} 
{1+\frac{(k+q)^2}{8m_{d^{\prime}\pi}\Delta E}}\right| 
\label{D0} 
\end{eqnarray} 

\noindent 
In the integrand of Eq.~(\ref{D0}) we use the definitions
$\Delta E=m_{d^{\prime}}+m_{\pi}-E$ and
$m_{d^{\prime}\pi}=m_{d^{\prime}}m_{\pi}/(m_{d^{\prime}}+m_{\pi})$. 
It should be stressed that the approximations $a_{\scriptscriptstyle L=0}$ 
and $\omega_{\pi}\approx\omega_{\mbox{\small nrel}}$ do not 
play any significant role in integration over $\cos\theta={\b k}\cdot
{\b q}/kq$ in Eq.~(\ref{D}) but without these approximations the explicit 
form of the integrand in Eq.~(\ref{D0}) becomes very complicated, and thus a 
qualitative consideration of the behavior of the amplitude based on the
decomposition~(\ref{DD}) would be impossible. However, starting from 
Eq.~(\ref{D}) one can obtain easily an important result: in 
the limit $k\to 0$ the integral in Eq.~(\ref{D}) goes to a non-vanishing 
value 
\begin{eqnarray} 
D^L(E,0)=4\pi\int\limits_0^{\infty} 
\frac{(1+\frac{5}{96}a_{\scriptscriptstyle L}q^2b^2)
\omega_{\pi}(q/2)e^{-q^2B^2}} 
{(m_{d^{\prime}}+\omega_{\pi}(q/2)+q^2/8m_{d^{\prime}}-E)\,
(m_{d_0}+2\omega_{\pi}(q/2)-E)}\,\,q^2dq, 
\label{D00} 
\end{eqnarray} 
i.e. the proposed mechanism of $\sigma - 6q$ coupling does not vanish 
at low $\sigma$-meson momenta. This should be very important for $\sigma$-meson  
effects in the $NN$ system and in nuclei. 
  
\subsubsection{Separable form of the effective $NN$ interaction} 
 
In accordance with the diagram in 
FIG.~4 the contribution of virtual $d_0+\sigma$ states to the $NN$ interaction 
in the $S$ and $D$ partial waves is defined by the matrix elements:   
\begin{eqnarray} 
A^{L^{\prime}L}_{NN\to d_0+\sigma\to NN}= 
\int d^3r^{\prime}d^3r{\Psi^{L^{\prime}}_E}^*({\b r^{\prime}}) 
V^{L^{\prime}L}_E({\b r^{\prime}},{\b r})\Psi^L_E({\b r})\,, 
\label{ALL} 
\end{eqnarray} 
where $V^{L^{\prime}L}_E({\b r^{\prime}},{\b r})$ is the separable  
amplitude of the virtual transition: 
\begin{eqnarray} 
V^{L^{\prime}L}_E({\b r^{\prime}},{\b r})=\int d^3k 
\Omega^{L^{\prime*}}_{NN\to d_0+\sigma}(E;{\b r^{\prime}},{\b k}) 
G^{(0)}_{d_0\sigma}(E;{\b k}) 
{\Omega^{L}_{NN\to d_0+\sigma}}(E;{\b r},{\b k}) 
\label{VLL} 
\end{eqnarray} 
Using the simple form of Eq.~(\ref{om1}) for transition amplitudes 
$\Omega^{L=0()2}_{NN\to d_0+\sigma}$ and the free Green function in the 
form: 
$$G^{(0)}_{d_0\sigma}=\left (E-m_{d_0}-m_{\sigma} 
-\frac{k^2}{2m_{d_0}}-\frac{k^2}{2m_{\sigma}}\right )^{-1}$$  
one can obtain eventually a matrix separable potential of special form 
\begin{equation} 
V_{NqN}(E)\equiv V^{L^{\prime}L}_E({\b r^{\prime}},{\b r}) 
=D_{NqN}(E)\left ( \begin{tabular}{cc} 
$g_0^2|2s({\b r^{\prime}})\rangle \langle 2s({\b r})|$ &  
$g_0g_2|2s({\b r^{\prime}})\rangle \langle 2d({\b r})|$ \\ 
$g_2g_0|2d({\b r^{\prime}})\rangle \langle 2s({\b r})|$ &  
$g_2^2|2d({\b r^{\prime}})\rangle \langle 2d({\b r})|$ 
\end{tabular} \right ),  
\label{sp} 
\end{equation} 
where the common energy-dependent factor $D_{NqN}$ has a complicated  
energy dependence 
\begin{equation} 
D_{NqN}(E)=  
\int k^2dk \frac{\exp\left (-\frac{5}{24}k^2b^2\right )|D(E,k)|^2}  
{E-m_{\sigma}-m_{d_0}-\frac{k^2}{2m_{\sigma d_0}}}. 
\label{am10} 
\end{equation} 
(we denote by  
$m_{\sigma d_0}$ the factor $m_{\sigma}m_{d_0}/(m_{\sigma}+m_{d_0})$ 
as usually) where the $D(E,k)$ is taken from Eq.~(\ref{DD}). 
Note that $D_{NqN}$ implies some $L$-dependence as follows from 
Eq.~(\ref{DD}). Starting from Eq.~(\ref{DD}) one can write down for $D_{NqN}$
a representation
\begin{eqnarray} 
D_{NqN}(E)=\tilde D_{NqN}(E)+a_La_{L^{\prime}}D^{\prime}_{NqN}(E),
\label{DDD} 
\end{eqnarray} 
where $a_L$ is defined by Eq.~(\ref{FL}). The contribution of the last
term of Eq.~(\ref{DDD}) to the matrix elements of Eq.~(\ref{sp}) can be taken
into account through a renormalization of values $g_Lg_L^{\prime}$ defined
in Eq.~(\ref{gL}). Thus, it is only a technical problem.
  
From the inspection of Eq.~(\ref{sp}) we find a very important result that the separable potential in Eq.~(\ref{sp}) gives
rise to a new type of short-range tensor force of the non-Yukawa type, which
originates basically from the coupling   between the input and output $L=0$
and $L=2$ form factors   $|2s({\b r^{\prime}})\rangle$ and $\langle 2d({\b
r})|$ through the intermediate $(0s)^6+2\pi $ states. In other words, this
new tensor   force comes from coupling of the cluster-like initial $NN$ channel
with the  mixed symmetry configuration $|s^4p^2[42]L\!=\!0,2\,\,
ST\!=\!10J^P=1^-\rangle $
and  the dressed bag intermediate states $(0s)^6+2\pi$. The above  specific
tensor term has a well expressed short-range character and this new 
interaction is very crucial to give a correct description of the mixing
parameter   $\varepsilon_1$ and the deuteron tensor structure with very soft
cut-off   factor $\Lambda_{\pi NN}$ in the OPE sector of our model. (We want
to remind to the reader that a correct description of $\varepsilon_1$ 
at higher energies and deuteron $D$-state admixture requires the
artificially   enhanced $\Lambda_{\pi NN}$ values in traditional OBE models,
see   Section~II.)

\section{A simple model} 
 In this section we demonstrate  that the mechanism of $NN$ interaction
developed in    preceding section is really able to describe $NN$ scattering in
a wide energy region.   For this aim we parameterize the basic potential
components  entering this model via a simple phenomenological form 
which includes the main features of the    above mechanism.  

Thus, the model
interaction consists of three terms: the orthogonalizing  potential   $V_{\rm
orth}$ providing the condition of orthogonality  between the proper  $NN$ 
channel and  the six-quark intermediate  bag in $S$- and $P$-waves,      the
one-pion-exchange potential  $V_{\rm OPE}$ with soft dipole truncation, and
the separable  term $V_{NqN}$ with an energy dependence described by a pole
(which  is the simplest approximation to a quark-induced interaction 
corresponding to the separable amplitude $V^{L'L}_E$~(\ref{VLL}) of the 
virtual transition $ NN \to (6q+2\pi ) \to NN$) as  illustrated by 
the graphs in FIG.~3 or FIG.~4:     

\begin{equation} 
 V_{NN}=V_{\rm orth}+V_{NqN} +V_{\rm OPE}, 
\label{mod1} 
\end{equation}  
\begin{equation} 
V_{\rm orth}=\lambda_0 |\varphi_0\rangle \langle \varphi_0 |, 
\qquad (\lambda_0 \to \infty) 
\label{mod2}  
\end{equation} 
\begin{equation} 
V_{\rm OPE}({\bf k})= \frac{f_{\pi}^2}{m^2}\, \frac{1}{{\bf k}^2+m^2} 
\left ( \frac{\Lambda^2-m^2}{\Lambda^2+{\bf k}^2}\right )^2  
({\b \sigma}_1{\bf k})({\b \sigma}_2{\bf k}) \frac{({\b \tau}_1{\b \tau}_2)}{3} 
\label{mod3} 
\end{equation} 

The interaction $V_{NqN}$ for single channel case takes the form:   
\begin{equation} 
V_{NqN}=\frac{E_0^2}{E-E_0} \lambda|\varphi\rangle \langle \varphi|,  
\label {mod4} 
\end{equation}
while the term $V_{NqN}$  for coupled  channels is a $(2{\times}
2)$-matrix  (compare  with Eq.(\ref{sp})): 
\begin{equation} 
V_{NqN}=\frac{E_0^2}{E-E_0}\left (  
\begin{tabular}{cc} 
$\lambda_{11}|\varphi_1\rangle \langle \varphi_1| \qquad$ &  
$\lambda_{12}|\varphi_1\rangle \langle \varphi_2|$ \\ 
$\lambda_{21}|\varphi_2\rangle \langle \varphi_1| \qquad$ &  
$\lambda_{22}|\varphi_2\rangle \langle \varphi_2|$ 
\end{tabular} \right ),  
\label{mod5} 
\end{equation} 
where one assumes $\lambda_{12}=\lambda_{21}$. For all form factors in 
eqs.(\ref{mod2}),(\ref{mod4}),(\ref{mod5}) we use the simple  Gaussian form
with one scale parameter $r_0$\footnote{ It is still in  accordance with our
general algebraic multiquark formalism due to an  appearance of the
additional orthogonality condition (see the  respective orthogonalizing
potential (\ref{mod2})).}: 
\begin{equation} 
\varphi_i(r)=
 Nr^{L_i+1} \exp \left (-\frac{1}{2}\left (\frac{r}{r_0}\right )^2  \right ) 
\label{mod6} 
\end{equation} 
 In the calculations the averaged pion mass
$m=(m_{\pi_0}+2m_{\pi_{\pm}})/3$,   the averaged value of pion-nucleon
coupling constant  $f^2_{\pi}/(4\pi ) = 0.075$, and a soft cut-off parameter 
with  values $\Lambda=\Lambda_{\rm dipole} = 0.50\div 0.75 $~GeV have been used.  

The results of the fits of the model parameters $\lambda_k $ (or
$\lambda_{jk}$), $r_0$ and $E_0$ to the $NN$ phase shift 
analysis data are displayed on FIGS.~6-8. It is quite evident this  simple
model  describes the $NN$ low partial waves up to   $E_{\rm lab}=600$~MeV
very well. The model phase shifts and the mixing parameter $\varepsilon_1$ 
are compared  in Figs.~6-8 with data of
a recent phase shift analysis (SAID,  solution SP99~\cite{SAID97}).  There are
three fitting  parameters for each partial wave: $\lambda$ ($\lambda_k $ or
$\lambda_{jk}$ for coupled channels), $r_0$ and  $E_0$.
The parameters of the projection operators (\ref{mod2})   ($r_0$ for $V_{\rm
orth}$) are taken from our preceding   work~\cite{PRC99} where the deep local
attractive potential   (Moscow potential) have been constructed as an
effective $NN$   one-component potential.   The parameter $E_0$ corresponds
to the sum of the six-quark bag energy and the effective $\sigma$-meson mass inside
the six-quark bag. Its value is  taken here in the range $600\div 1000$~MeV. 
In accordance to our suggestions,
it should be the  same for all partial waves with definite parity. We found
that  the  results  depend on $E_0$ only weakly.    All parameters found for
$S$-, $P$- and $D$-waves are given in Table~\ref{table1}. 

It is highly instructive to
compare the present simple model based on the suggested new mechanism for 
$NN$ interaction with the well known phenomenological separable
potential~\cite{Graz} (so called Graz potential) which fits the same phase
shifts until $E_{\rm lab}=500$~MeV. The reader can find from the comparison
that the number of free parameters in the Graz potential exceeds very much those
for our simple model (\ref{mod1}) whereas the energy range 
is smaller and the quality of the fit is worse for the Graz model. Thus our simple
model describes $NN$ data more adequately than the Yamaguchi-type
phenomenological model.

 Moreover, it was very  surprising to find out that such a simple model gives a very good
description   for $^1S_0$ phase shifts even up to $E_{\rm lab}=1200$~MeV
(see  FIG.~8)(at higher energies  the $np$ phase shift analysis (PSA) is 
absent).  

 We want to discuss here especially the description  of phase shifts in  
triplet coupled channels $^3S_1 -  {}^3D_1$.  The crucial point is the behavior
of the mixing parameter  $\varepsilon_1$  with increasing energy.  Without
the separable (quark-bag induced) mixing potential (i.e. at $\lambda_{12}=0$) the  
behavior of $\varepsilon_1$ is correct only at very low energies, but  
is in strong contradiction with the  PSA  at energies higher than  50 MeV
(see the dashed  line in FIG.~7.)  The increase of the  truncation parameter
$\Lambda$ up to values 0.8 GeV does not help to  get a  better agreement with
the data, but on the contrary, destroys  the good description at low  energies
(the dotted line in the  Figure). Introducing the quark-bag induced mixing
($\lambda_{12}\ne  0 $ in Eq.(\ref{mod5})) allows us to reproduce  the
behavior of  $\varepsilon_1$ (and $^3S_1 - {}^3D_1$ phase shifts as well)  
with a reasonable accuracy until the energy as high as $E_{\rm lab}\sim 
600$~MeV, but only for sufficiently small values of $\Lambda_{\pi  NN}$. 
The best fit for the $\varepsilon_1$ mixing parameter is shown on FIG.~7 
(by solid line) with the potential parameter values given in
Table~\ref{table1} where 
the value of $\Lambda_{\pi NN}=0.5936$~GeV.  

In this case the condition: 
\begin{equation} 
\lambda_{12}^2=\lambda_{11}\lambda_{22} 
\label{mod7} 
\end{equation} 
is satisfied with high accuracy. Just this  condition follows from  our
assumption that the quark-bag induced  $S-D$ mixing arises due to coupling 
of the $NN$ channel with $L=0,2$ to a single  $S$-wave six-quark state 
$|s^6+2\pi \rangle$ (see Eq.(\ref{sp})). The  increase of  
$\Lambda_{\pi NN}$ up to a value 0.8~GeV results in the violation of  
condition (\ref{mod7}) and in a significant deterioration of 
the description of $\varepsilon_1$ (see the dot-dashed   line in the  FIG.~7).
The other phase shifts ($^3S_1$ and $^3D_1$) are reproduced   for all four
variants with the same good accuracy, so that we  present in FIG.~6   the
results for one variant only. 

 Thus  we can deduce from the results of our simple model presented in  this 
Section, that the model is able to describe all  phase  shifts in low partial
waves $L=0\div 2$ in a rather large energy  interval   $0 \div 600$~MeV. This
good description and the comparison with the phenomenological Graz model 
seems support  strongly the new  dressed bag mechanism for
the intermediate-range  interaction suggested in the work. Moreover in  the
next Section we  will show a tight interrelation existing between the
current  microscopic model and various phenomenological models of the $NN$
interaction  proposed  in earlier years. 

\section{Relationships of the suggested mechanism to other interaction 
models}  

In this section we will discuss the interrelations of the  new $NN$ 
mechanism suggested   in this work with other models  proposed in previous years
and elucidate   the microscopic grounds for  some of them. 

\subsection{Relationship to the Moscow potential model} 

 While the symmetry background of the Moscow potential 
models~\cite{PRC99,Kuk90,Kuk90Jap,Progr92,Kuk98,Moscow} is rather   
similar to the
present model the underlying mechanism and the particular  realization are
very different. In the above potential models one starts  with  a subdivision
of the possible spatial (permutational) six quark  symmetries of  the total
wavefunction into two types of different  physical nature: $\Psi_{\rm 
bag}\left ((0s)^6[6]\right )+  \Psi_{NN}\left ((0s)^4(1p)^2[42]\right )$  
which are mutually  orthogonal  to each other. Then by excluding the bag-like
components  from the proper  $NN$ channel one arrives at an effective 
interaction Hamiltonian in the $NN$  channel~\cite{Kuk90,KuFaes98} with  an
additional orthogonality condition constraint: 

\begin{mathletters} 
\begin{equation} 
\left (T_R+V_{\rm ME}+ 10\frac{|f\rangle \langle f|}{E-E_{0}}\right )\tilde{\chi}=  E\tilde{\chi} 
\label{inter1a} 
\end{equation} 
\begin{equation} 
\langle g|\tilde{\chi}\rangle =0. 
\label{inter1b} 
\end{equation} 
\label{inter1} 
\end{mathletters} 
$\tilde{\chi}(R)$ is the wavefunction of the relative motion in the $NN$
channel   which is  renormalized  through the overlap kernel ${\cal N}({\bf
R,  R'})$ to  have a probabilistic meaning~\cite{KuFaes98}. Here $V_{\rm ME}$
is the  sum of  conventional meson-exchange potentials truncated at the 
proper (i.e.  soft) values of $\Lambda_{mNN}$,  while the form  factor
$f({\bf R})$ in the separable term of Eq.~(\ref{inter1a}): 
$$ 
 \langle {\bf R}|f\rangle  \equiv f({\bf R})=\langle \psi_{6q}|H|\psi_{N}\psi_{N}\rangle  
$$ 
 is the matrix element which couples the six-quark and $NN$  channels, and
the  function $g({\bf R})$ in the orthogonality  condition (\ref{inter1b})
is  taken as:  
 $$\langle {\bf R}|g\rangle  \equiv 
g({\bf R})=\langle \psi_{6q}|\psi_{N}\psi_{N}\rangle . 
 $$ 

 Then in the initial version of the one-channel Moscow $NN$
potential~\cite{Kuk90,Moscow,LIYAF86} one  replaces  both the separable term in
(\ref{inter1a}) and the  orthogonality  condition~(\ref{inter1b}) by one deep
local potential  where the  deep-lying bound states (which are considered as 
"forbidden" states in  the model) serve to provide the orthogonality 
condition~(\ref{inter1b})  due to the hermicity of the underlying  potential. 

 Thus, the previous Moscow $NN$ potential model is  in essence a  phase-shift
equivalent local effective potential to a highly non-local and 
energy-dependent model~(\ref{inter1}). Our  next step was the  generalized
orthogonality-condition  model~\cite{PRC99} where we still retained the deep 
local potential  but disconnected really the bound state wavefunction  in
the  potential from the orthogonality condition. Thus from this   point  of
view the  above $NN$ model can be considered as a generalized  orthogonality
condition model initially proposed by Saito in  nuclear  cluster
physics~\cite{Saito} as early as 1969. And very  similar as in the cluster
model, the deep attractive well of the  one-channel Moscow $NN$  potential
represents a local phase-shift  equivalent potential for a nonlocal and
energy-dependent interaction  term in Eq.(~\ref{inter1a}) together  with the
orthogonality  condition constraint~(\ref{inter1b}). As a result  of the 
constraint, the $NN$ phase shifts in low partial waves ($S$ and  $P$) display
a behavior similar to phase shifts for a local repulsive core  
potentials~\cite{Re68}. In
other words, the orthogonality condition is   equivalent in some sense to a
repulsive core, but only on the energy  shell. The
orthogonality condition results in a  stationary (with respect to the energy  
variation)
short-range node in the $NN$ wavefunction of the relative motion rather  
than to   a
strong damping  of the latter near the origin. Moreover the node position  ($r_n
\simeq  0.6$~fm) coincides very nearly to the radius of repulsive  core in 
the traditional force models like RSC~\cite{Re68} etc. 

 It is also  instructive to know that the node position is very near to the
size of the 
repulsive core only if the quark radius of the nucleon  $r_q$  is about
0.6~fm. If one assumes a smaller quark radius of the  nucleon 
(around $\bar r_q \simeq 0.35$~fm), as in  some  modern models for
baryons~\cite{Gloz96}, the node position is shifted inward and
one needs extra  repulsive  terms~\cite{KuFaes98} (in addition
to the node) to  describe adequately the $NN$  phase shifts. In this way the 
short-range stationary node in wavefunction of the relative  motion 
replaces a
big portion of the repulsive core and thus  the
coupling constant for $\omega$-meson  exchange may be reduced safely to the
moderate values  $g^2_{\omega  NN}/4\pi \simeq 5$ dictated by $SU(3)$
symmetry.  Thus the new  dressed bag model presented in the work gives a
microscopic  quark-meson realization for the previous Moscow-type $NN$
models.    

\subsection{Interrelation with the Simonov's QCB and other hybrid 
models} 
  
 The total wavefunction of $NN$ system, according to  the Simonov's 
quark-compound bag (QCB) model, is composed,  similarly to our basic 
assumption (1) from two components of  different nature: the quark  compound
bag part at small distances  $r\le R_0$ and the proper  cluster-like $NN$
component in peripheral  region $r > R_0$ with $R_0$   being the  matching
radius between  the two components. Similar arguments can be  presented
also for  other hybrid models (e.g. for Kisslinger et al.  
model~\cite{Kis}).  Then, analogously  to our  formal
derivation~\cite{KuFaes98}, the  bag-like component is eliminated in the QCB
approach and one arrives at an effective  one-channel Schr\"odinger
equation for the $NN$ component analogous  to~(\ref{inter1a}) where the
transition form factor $f({\bf R})$ in the  QCB  model is chosen as
a $\delta$-function centered at the transition  radius $R_0$  plus energy dependent
terms\cite{Sim}. However, in the QCB  approach in contrast to our  model, two
basic components, i.e.  $\Psi_{6q}$ and $\Psi_{NN}$ are taken   to be
nonorthogonal to each  other. Thus, the very important 
constraint~(\ref{inter1b}) is  absent in the  QCB-model\footnote{This
nonorthogonality of two  basic components in QCB   leads to an appearance of
some ghost state  in infinity which can be  considered as an analog of deeply
bound  "forbidden" states in our  approach.}. However, when the two channels 
are orthogonalized in the QCB  approach the  scattering wavefunctions  in $NN$
channel acquire a short-range node  rather similar to our  case but with 
a violation of the continuity at the matching radius $R_0$. Another important feature
which distinguishes   our  two-component approach from the QCB and other
hybrid models is the fact that the both components in our approach are treated
in Hilbert space  while in the hybrid models they are taken in 
configurational space. 

 But the main difference of the QCB approach to our current model  is  the
fact that the QCB is, in essence, a phenomenological model (based  on the $P$-matrix
formalism) which does not consider any microscopical  or  field-theoretical
aspects. However the fact that, starting  with  absolutely independent
arguments (in fact we started more than  two decades   ago with the old 
phenomenological Moscow type $NN$  potential\cite{ru73}), we arrived at a  
model which, in its formal  aspects, has many similarities with QCB,   shows
that both models  reflect the underlying true physical picture  rather
adequately. 

 \subsection{Interrelation with the separable Tabakin 
potential} 

 There are also very interesting connections between  our approach and  the
 Tabakin potential.  More than 30 years  ago Tabakin, to facilitate
drastically the Faddeev   few-nucleon  calculations, proposed~\cite{Tab68}
the   phenomenological one-term  separable potential "with repulsion and 
attraction". The  characteristic feature of the Tabakin potential which 
distinguishes  it from many separable models proposed at that time  is   an 
oscillating behavior of the potential form factor $g(p)$ in  $S$-waves: 
\begin{mathletters} 
\begin{equation} 
 V_{\rm T}(p,p')=\lambda g_{\rm T}(p)g_{\rm T}(p') 
\label{inter2a} 
\end{equation} 
with 
\begin{equation} 
 g_{\rm T}(p)=\frac{p^2-p_0^2}{(p^2+\beta^2)^2} 
\label{inter2b} 
\end{equation} 
\label{inter2} 
\end{mathletters} 
 The  change in sign  of $g_{\rm T}(p)$ at $p=p_0$ was able to produce  a 
respective change in the sign of $S$-wave phase shift at  $E_{\rm  lab}\simeq
300 $~MeV due to appearance of the so called  continuum bound  state (CBS).
In other words, the one-term Tabakin  potential was able to describe  both
low-energy attraction and  high-energy repulsion in $S$-wave of the $NN$
interaction\footnote{Almost simultaneously with the Tabakin  work we 
suggested\cite{Kuk70} very similar separable potentials to describe 
cluster-cluster interaction for the systems such as  $^4$He-${}^4$He, 
$^4$He-$d$ etc. where all lowest partial phase  shifts also change their 
sign (from positive to negative) at rather  low energies.}. 

 At that time the success of the Tabakin  potential was considered  as
somewhat "accidental" and puzzling.  However about a decade ago 
Nakaishi-Maeda has  demonstrated\cite{Nak86} that the Tabakin potential can
be  considered in a very good approximation as a first term in the 
unitary-pole expansion of $t$-matrix for the deep local Moscow $NN$ 
potential while the scattering wavefunctions for both models display the
short-distance stationary nodes (at $r_n \simeq 0.6$~fm) in very 
similar ways. Moreover, it has been  shown~\cite{Nak86} that the 
continuum bound state in the Tabakin potential has the  
energy $E\simeq 300$MeV (in the lab. 
system) and is very similar in its structure to the  "forbidden" bound state in the  initial
version of the Moscow  potential. 

 Here we want to demonstrate that the analogy  between the new quark-meson 
mechanism suggested in the present  paper and the old Tabakin potential  goes
much further. In fact, the  overlap factors (\ref{prj} ) between  three-quark
nucleon clusters  and six-quark  configurations $|s^4p^2[42], L=0,ST\rangle $
and  $|s^6[6], L=0,ST\rangle $ lead inevitably to the nodal $2S$-type 
relative motion  form factors in our separable potential term  (\ref{mod2} ).
In the momentum representation this  form factor  behaves like: 
\begin{equation} 
g_{2S}(p)=N_{2S}(p^2-p_0^2)\exp(-\frac{3p^2}{4 p_0^2}) 
\label{inter3} 
\end{equation} 
 which has the same nodal  character with the same node position $p_0^2$  as
the Tabakin's form  factor~(\ref{inter2b}). The only difference is the 
power-like  truncation factor $(p^2+\beta^2)^{-2}$ in~(\ref{inter2b}) is 
replaced in our case with the Gaussian form in~(\ref{inter3}) due  to 
the use of six-quark shell-model basis in our  calculations.  Making use
of the $2s$-type form factor~(\ref{inter3}) will  project out all the admixture of
nodeless  $0s$-components in $NN$ scattering  wavefunctions giving by this
way  a stable short-range node in the $S$-wave  at $r_0\simeq 0.6$~fm. Thus, 
the use of the oscillating $2s$-type  form factors replaces, in a  good
approximation, our orthogonality  condition constraints  (\ref{pr2}) and
(\ref{inter1b}), resulting, in essence, in  almost  the same scattering
wavefunctions. Therefore one can conclude that   the Tabakin one-term
potential agrees qualitatively with our new mechanism which dictates just
a $2s$-type oscillating character of the  potential form   factors or
alternatively the necessity for the additional orthogonality constraint. 
This gives
a quark microscopic  interpretation for the success of the old
phenomenological Tabakin  potential. From here one can conclude that there
are many  completely  independent arguments in favor of our new mechanism of 
interaction  suggested in this work. 

\section{Conclusion} 

 In this paper we presented a critique of the  conventional 
meson-exchange models of nuclear forces at  intermediate and short ranges.
We provided many arguments  demonstrating clearly the inner
inconsistencies and contradictions  in modern OBE models for the short-range
part of the interaction. There  are also several observations in few-nucleon
systems showing  clearly that one cannot  explain quantitatively and
consistently  many $3N$ and $4N$ experimental data with the
existing  $NN$ models. 

To find an alternative picture of  the $NN$ interaction we 
exploited the successful quark  motivated  semi-phenomenological
models, viz. the  Moscow~\cite{PRC99,Moscow,KuFaes98} and  T\"ubingen
microscopic  quark approaches~\cite{Buch88,Fae83,Buch89}, to develop them 
further. In this way we suggested in the
present paper  some  new mechanisms for the intermediate- and short-range $NN$ 
interaction. These  mechanisms are distinguished from the  traditional Yukawa
concept for the  meson exchange in the $t$-channel. We introduce a concept of
the dressed  symmetric six-quark bag in  the intermediate state with 
$s$-channel  propagation. In a tight  connection to this mechanism we
proposed also  a new interpretation  of the scalar-isoscalar $\sigma$-meson
as a quasiparticle, i.e. the particle can mainly only exist inside hadrons 
rather than as a real  resonance in free space. In this respect the 
$\sigma$-mode  should be treated differently as compared 
to other mesons like  $\rho$, $\omega$ etc. The new interaction mechanism proposed here  has
been shown to lead to separable energy-dependent $s$-channel   resonance-like
interaction terms plus a term with a projection operator (in  lowest partial waves)
resulting from a constraint for a orthogonality condition. 

In its final form the
proposed  interaction depends only on a few fundamental constants
(quark-meson  or diquark-meson  coupling constants and the intermediate meson
masses) so  that  eventually the total $NN$ force can be parametrized
by a few free parameters only. However, at the present stage 
we prefer to  employ the derived form of the interaction to
build a  simple model whose  main goal is to illustrate how well the 
suggested mechanism can work.     We found that by adjusting only  three 
parameters of the model in each   partial wave it is possible to
describe excellently all lowest $NN$ phase shifts  in a large energy interval
$0 \div 600$~MeV, and $S$-waves even until 1200~MeV in the lab. frame. This 
gives some strong
evidence that the suggested new  microscopic mechanism of $s$-channel
"dressed" symmetric bag  should  work adequately. 

 The proposed interaction model has  been demonstrated to give a natural 
microscopic background for  previous phenomenological interaction models 
like the Moscow $NN$  potential, the Tabakin separable potential "with 
attraction and  repulsion" and also the QCB model by Simonov and other 
hybrid  models. Thus  it gives also very important bridges between at first
glance absolutely  disconnected models developed previously.  In fact,
without the present model it  would be  extremely hard or impossible to
establish any correlations  between,  say, the Simonov QCB model and one-term
Tabakin potential. 

Another important result of the present model could be a possible  resolution
of a long-standing puzzle about the weak vector-meson  contribution  in baryon
spectra and a strong spin-orbital splitting  (due to the vector  meson
contribution) in the $NN$ interaction. If one assumes a significant  quark-quark
force due to vector-meson (or one gluon) exchange the vector coupling  will result
immediately  also in  strong spin-orbit splitting in baryon  spectra. Very
recently  Glozman and Riska~\cite{Gloz96} described the 
absence of spin-orbital splitting in negative  parity excited  baryon
states in a model for the $qq$  interaction 
 mediated essentially by Goldstone-boson exchange.  However  this model
fails to explain the strong  spin-orbit splitting
in the $NN$ sector. Our explanation of  the puzzle is  based on the fact that
there is no significant  vector-meson  contribution into $qq$ forces (in 
$t$-channel)~\cite{Gloz96} but there is an important  contribution  of 
vector mesons in the dressing of the symmetric  six-quark bag  leading
thereby to strong spin-orbital effects in the $NN$  interaction  mediated by the
"dressed" bag. 

 Moreover, the proposed model  will lead to the appearance of strong $3N$ 
and $4N$ forces  mediated by $2\pi$ and $\rho$ exchanges (see e.g. 
$3N$-forces  graphs in FIG.~9).\\ 

\noindent  
{\mbox{ }\hfill\epsfxsize=0.7\textwidth\epsfbox{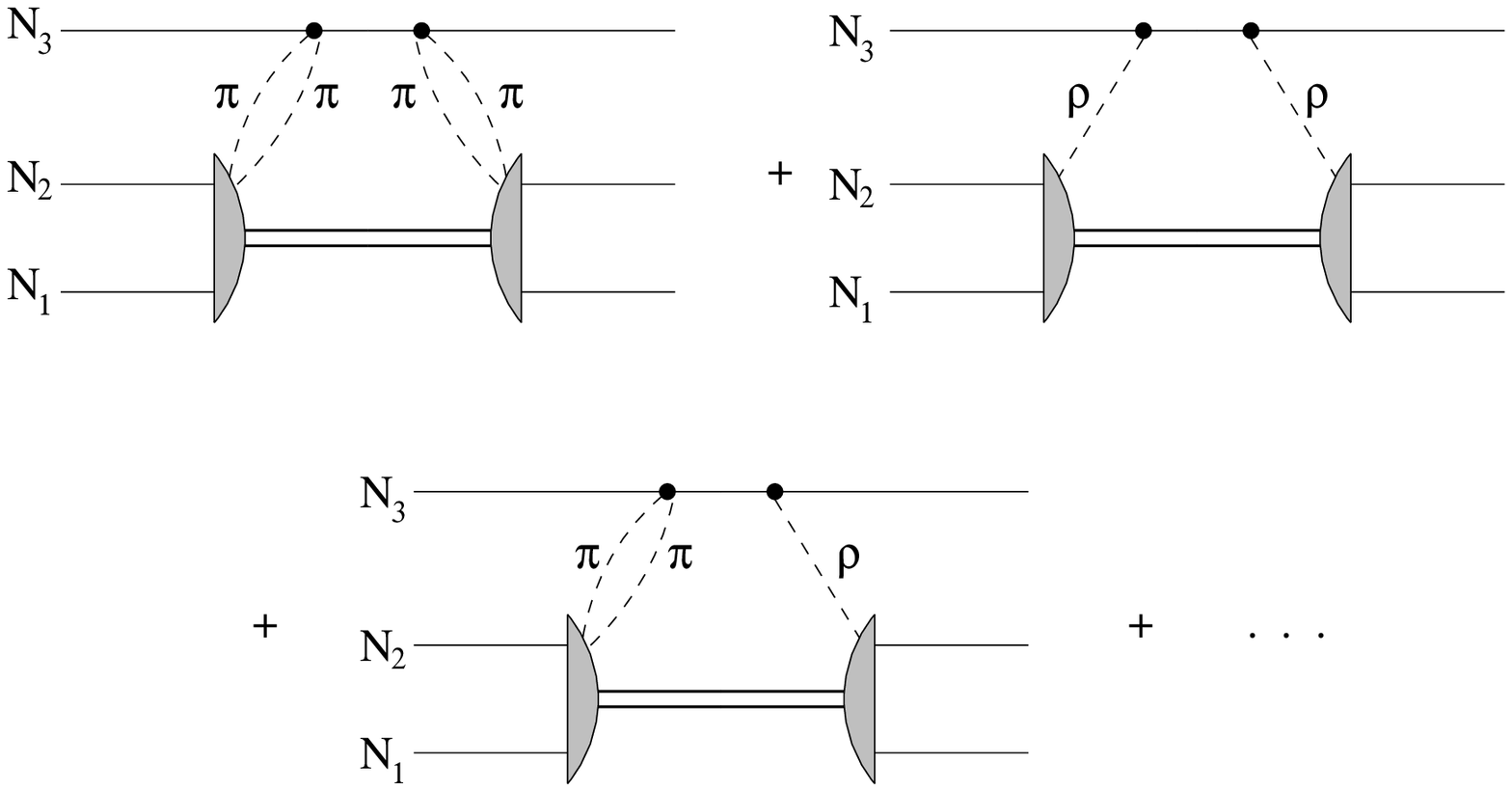}\hfill  
\mbox{ }} 
  
{\bf FIG.~9.} {\small Some graphs illustrating the new-type of $3N$ 
forces. }\\

 It is easy to see that the  new $3N$-forces include both central and 
spin-orbit components.  Such a spin-orbit $3N$ force is extremely  desirable
to explain  the low energy puzzle of the analyzing power  $A_y$ in $Nd$ 
scattering~\cite{FB98,Glo96}  and also the behavior of $A_y$ in the $3N$
system at higher energies $  E_N\simeq 250 \div 350 $~MeV at backward
angles~\cite{Mey99}. The  central  components of the $3N$ and $4N$ forces are
expected to be  strongly  attractive  and thus they must contribute to $3N$-
and  $4N$-binding  energies possibly resolving hereby the very old puzzle 
with the binding  energies of the lightest nuclei.  Moreover these strong
contributions (as one can expect) of the above $3N$-  and $4N$-forces
mediated by the "$\sigma$-type" $2\pi$-exchange to the  nuclear binding in a
combination with strong relativistic effects predicted by  our
model~\cite{Kuk99,LIYAF86} can lead very naturally to the relativistic 
hadrodynamics (i.e. the Walecka model) where the sigma-field constitutes the 
main agent for nuclear binding. The  suggested new mechanism leads to a large
number of new contributions  for  many  nuclear physics observables like
enhanced Coulomb displacements  energies for isobar-analog
states~\cite{Kuk98,KuFaes98}, enhanced  spin-orbit splitting in  the nuclear
shell model, more  significant relativistic effects, a  serious
renormalization of  meson-exchange current contributions etc.,  etc. Future
studies  must  show to what degree such  expectations can be  justified. 

\centerline{\large \bf Acknowledgments.}  The authors express their deep
gratitude to many colleagues for  continuous encouraging.   Special
thanks go to Prof. Steven  Moszkowski for his permanent support  of our
studies. We thank also  Prof.~Mitja Rosina and   Dr.~Alfons Buchmann.  The
Russian authors  thank the  Russian Foundation for Basic Research (grant 
No.97-02-17265) and the  Deutsche Forschungsgemeinschaft
(grant No.  Fa-67/20-1) for partial financial  support. 


\appendix
\section{Details of quark-model calculations}
Here we consider some details for the quark-model calculations of the two-pion 
emission amplitude for the transition from the $^3S_1(^3D_1)\,\,\, NN$ state 
to the "dressed" six-quark bag $d_0+\pi+\pi$. We will demonstate here 
how to use the known 
fractional-parentage coefficient (f.p.c.) technique 
\cite{Obu79,Harv81,Kus91,Obu96,Obu99}
in calculations of the two-step prosses
$d_f\to d^{\prime\prime}+\pi\to d_0+\pi+\pi$ in the channel 
$ST\!=10\,\,J^P\!=1^-$. First we consider the two-pion emission in the
two-quark subsystem "56" (where "5" and "6" are quark numbers in the six-quark 
system "123456"). We start from the 2s(2d) h.o. state of the 6-th quark in the
$d_f$-state (see Sects. IV, Eqs. (\ref{df}) and (\ref{d1})-(\ref{d2})) 
which, after S-wave pion emission, 
goes to the 1p h.o state in the 56-subsystem of the 
intermediate $d^{\prime\prime}$ configuration. At the next step, the 5-th
quark of the 56-subsystem emits another S-wave pion and the 
intermediate $d^{\prime\prime}$
configuration goes to the final $d_0$ configuration in which the 
56-subsystem is in the 0s h.o. state. Therefore, we must take into 
consideration the following five non-vanishing elementary $q\to q+\pi$ 
transition amplitudes in the h.o. quark basis:

({\it i}) the two amplitudes 
\begin{eqnarray}
T^{(6)}_{2s\to 1p}(j_{56}\!=\!0)\equiv \langle 1p,s_{56}\!=\!1(j_{56}\!=\!0),
t_{56}\!=\!0|H^{(6)}_{\pi qq}|2s,s_{56}\!=\!0,t_{56}\!=\!1\rangle\nonumber\\
=i\,v\,\sqrt{\frac{2}{3}}\frac{\omega_{\pi}(k_6)}{2m_q\alpha}
(1/2||\sigma||1/2)\sqrt{\frac{1}{6}}
(1/2||\tau||1/2)F_2^{L\!=\!0}(k_6^2),\nonumber\\
T^{(5)}_{1p\to 0s}(j_{56}\!=\!0)\equiv \langle 0s,s_{56}\!=\!0,
t_{56}\!=\!1|H^{(5)}_{\pi qq}|1p,s_{56}\!=\!1(j_{56}\!=\!0),
t_{56}\!=\!0\rangle
\nonumber\\
=i\,v\,\frac{\omega_{\pi}(k_6)}{2m_q\alpha^{\prime}}
(1/2||\sigma||1/2)\sqrt{\frac{1}{6}}
(1/2||\tau||1/2)F_0(k_5^2)
\label{ap1}
\end{eqnarray}
should be taken if the total angular momentum of the 56-subsystem 
$j_{56}=0$ and

({\it ii}) the three amplitudes
\begin{eqnarray}
T^{(6)}_{2s\to 1p}(j_{56}\!=\!1)\equiv \langle 1p,s_{56}\!=\!0(j_{56}\!=\!1),
t_{56}\!=\!1|H^{(6)}_{\pi qq}|2s,s_{56}\!=\!1,t_{56}\!=\!0\rangle\nonumber\\
=-i\,v\,\frac{\sqrt{2}}{3}\frac{\omega_{\pi}(k_6)}{2m_q\alpha}
(1/2||\sigma||1/2)\sqrt{\frac{1}{6}}
(1/2||\tau||1/2)F_2^{L\!=\!0}(k_6^2),\nonumber\\
T^{(6)}_{2d\to 1p}(j_{56}\!=\!1)\equiv \langle 1p,s_{56}\!=\!0(j_{56}\!=\!1),
t_{56}\!=\!1|H^{(6)}_{\pi qq}|2d,s_{56}\!=\!1,t_{56}\!=\!0\rangle\nonumber\\
=-i\,v\,\frac{\sqrt{2}}{3}\frac{\omega_{\pi}(k_6)}{2m_q\alpha}
(1/2||\sigma||1/2)\sqrt{\frac{1}{6}}
(1/2||\tau||1/2)F_2^{L\!=\!2}(k_6^2),\nonumber\\
T^{(5)}_{1p\to 0s}(j_{56}\!=\!1)\equiv\langle 0s,s_{56}\!=\!1,(j_{56}\!=\!0),
t_{56}\!=\!0|H^{(5)}_{\pi qq}|1p,s_{56}\!=\!0,t_{56}\!=\!1\rangle
\nonumber\\
=-i\,v\,\sqrt{\frac{1}{3}}\frac{\omega_{\pi}(k_6)}{2m_q\alpha^{\prime}}
(1/2||\sigma||1/2)\sqrt{\frac{1}{6}}
(1/2||\tau||1/2)F_0(k_5^2)
\label{ap2}
\end{eqnarray}
should be taken in the case of $j_{56}=1$.

To simplify matters we use shorthand notations $T^{(6)}_{2s\to 1p}
(j_{56}\!=\!1)$, $T^{(5)}_{1p\to 0s}(j_{56}\!=\!0)$,\dots, etc. for 
the elementary amplitudes
and omit spin, isospin and angular momentum projections (omitting the
summation over these quantum numbers in the following expressions).
The other shorthand notations are
$\alpha=\sqrt{\frac{6}{5}}\,b,\quad\alpha^{\prime}=-\sqrt{\frac{5}{2}}\alpha$, 
where $b$ is the scale parameter (r.m.s. radius) of the h.o basis functions, and
\begin{eqnarray}
F_0(k_5^2)=\exp(-\frac{5}{24}k_5^2b^2),\qquad 
F_2^{L}(k_6^2)=(1+\frac{5}{24}a_{\scriptscriptstyle L}k_6^2b^2)\,
\exp(-\frac{5}{24}k_6^2b^2),
\label{ap3}
\end{eqnarray}
where $a_{\scriptscriptstyle L}=\frac{1}{3}$,
if L=0 and $a_{\scriptscriptstyle L}=
-\frac{2}{3}$, if L=2.
The functions in (\ref{ap3}) provide the $k^2$-dependence of the form factors
in the $\pi d^{\prime\prime}d_f$- and $\pi d_0d^{\prime\prime}$-vertices
(see Eq. (\ref{fF}) in Sect. IV)
\begin{eqnarray}
F_{\pi d_0d^{\prime\prime}}(k_5^2)=F_0(k_5^2),\qquad 
F^L_{\pi d^{\prime\prime}d_f}(k_6^2)=F_2^{L}(k_6^2)
\label{ap4}
\end{eqnarray}
The reduced matrix elements $(1/2||\sigma||1/2)$ and $(1/2||\tau||1/2)$
of the spin(isospin)-flip operators (i.e. $\sigma$- and $\tau$-matrices in
the vertex operators $H^{(6)}_{\pi qq}$ and $H^{(5)}_{\pi qq}$) are defined 
here in accordance with
the Wigner-Ekkart theorem. Standard calculations give:
\begin{eqnarray}
(1/2||\sigma||1/2)=(1/2||\tau||1/2)=-\sqrt{6}
\label{ap5}
\end{eqnarray}

Recall that for the desired amplitude we use the 
parametrization of Eq. (25) of Sect. 4   
\begin{eqnarray}
15\,\langle d_0|H^{(5)}_{\pi qq}(k_5)|d^{\prime\prime}\rangle\,
\langle d^{\prime\prime}|H^{(6)}_{\pi qq}(k_6)|d_f\rangle \nonumber\\
=v^2\,\frac{\omega_{\pi}(k_5)\omega_{\pi}(k_6)}{m_q^2b^2}\,
f_{\pi d_0d^{\prime\prime}}f^{L}_{\pi d^{\prime\prime}d_f}\,
F_{\pi d_0d^{\prime\prime}}(k_5^2)\,F^L_{\pi d^{\prime\prime}d_f}(k_6^2).
\label{ap6}
\end{eqnarray}
Now one can calculate "the coupling constants"
$f_{\pi d^{\prime\prime}d_f}$ and $f_{\pi d_0d^{\prime\prime}}$ 
of this parametrization starting from the elementary amplitudes of 
Eqs.~(\ref{ap1})-(\ref{ap2}). For this purpose one can apply f.p.c.'s 
to separate the two-quark subsystem "56" 
from the six-quark configurations $d_f$, $d^{\prime\prime}$ and
$d_0$ for all possible color-, spin-, isospin- and coordinate states of
the quark pair ($[f_{56}]_{\scriptscriptstyle C}$= $[2]$,$[1^2]$, 
$s_{56}=$0,1, $t_{56}$=0,1, $j_{56}$=0,1 for 2s-, 2d-, 1p-, and 0s-radial 
and orbital states). Recall that the f.p.c. technique implies summation
over all possible states of the separated two-quark subsystem instead of
summation over all numbers of quarks in the interaction operator. 
This scheme is particularly handy for application 
of the group-theoretical algebraic methods.

We use the invariants (i.e. Young schemes 
$[f_{\scriptscriptstyle C}]$, $[f_{\scriptscriptstyle S}]$, 
$[f_{\scriptscriptstyle CS}]$, $[f_{\scriptscriptstyle T}]$, 
$[f_{\scriptscriptstyle CST}]$ and $[f_{\scriptscriptstyle X}]$) of the 
chain of symmetry groups (see, e.g. \cite{Obu96,Obu99})
\begin{eqnarray}
SU(12)_{CST}\supset SU(6)_{CS}{\times} SU(2)_T\supset
SU(3)_C{\times} SU(2)_S{\times} SU(2)_T,\nonumber\\
SU(24)_{XCST}\supset SU(12)_{CST}{\times} SU(2)_X
\label{ap7}
\end{eqnarray}
for classification of six-quark, four-quark and two-quark states in the
systems "123456", "1234" and "56" respectively. The f.p.c. for separation 
out of the
pair "56" in the total $XCST$ space $\Gamma_{XCST}(q^6\to q^4{\times} q^2)$ 
is a product of "scalar factors" of the Clebsch-Gordan coefficients
of groups $SU(6)_{CS}\supset SU(3)_C{\times} SU(2)_S$, 
$SU(12)_{CST}\supset SU(6)_{CS}{\times} SU(2)_T$ and 
$SU(24)_{XCST}\supset SU(2)_X{\times} SU(12)_{CST}$
taken from the reduction chain of Eq.~(\ref{ap7}) 
($\Gamma_{\scriptscriptstyle C\cdot S}$, 
$\Gamma_{\scriptscriptstyle CS\cdot T}$ and 
$\Gamma_{\scriptscriptstyle X\cdot CST}$) and "orbital"
f.p.c.'s $\Gamma_{\scriptscriptstyle X}$ of translationally-invariant 
shell model (TISM)
\begin{eqnarray}
\Gamma_{\scriptscriptstyle XCST}(q^6\to q^4{\times} q^2)=
\Gamma_{\scriptscriptstyle C\cdot S}
\Gamma_{\scriptscriptstyle CS\cdot T}
\Gamma_{\scriptscriptstyle X\cdot CST}\Gamma_X
\label{ap8}
\end{eqnarray}
The following extended notations for non-trivial scalar factors
$\Gamma_{C\cdot S}$ and $\Gamma_{CS\cdot T}$ are used here (see 
e.g.~\cite{Obu96}):
\begin{eqnarray}
&\Gamma_{\scriptscriptstyle C\cdot S}^{S\!=\!1}
([f_{\scriptscriptstyle CS}]([2^2]_{\scriptscriptstyle CS} 
{\times}[2]_{\scriptscriptstyle CS}),s_{56}\!=\!1)
\equiv\left(\begin{array}{cc}
[2^3]_{\scriptscriptstyle C}&[42]_{\scriptscriptstyle S}\\
([2^2]_{\scriptscriptstyle C}{\times}[2]_{\scriptscriptstyle C})&
([2^2]_{\scriptscriptstyle S}
{\times}[2]_{\scriptscriptstyle S})
\end{array}
\Biggl|\Biggr|
\begin{array}{c}
[f_{\scriptscriptstyle CS}]\\
([2^2]_{\scriptscriptstyle CS}{\times}[2]_{\scriptscriptstyle CS})
\end{array}\right),&\nonumber\\
&\Gamma_{\scriptscriptstyle C\cdot S}^{S\!=\!1}
([f_{\scriptscriptstyle CS}]([2^2]_{\scriptscriptstyle CS}
{\times}[2]_{\scriptscriptstyle CS}),s_{56}\!=\!0)
\equiv\left(\begin{array}{cc}
[2^3]_{\scriptscriptstyle C}&[42]_{\scriptscriptstyle S}\\
([21^2]_{\scriptscriptstyle C}{\times}[1^2]_{\scriptscriptstyle C})&
([31]_{\scriptscriptstyle S}
{\times}[1^2]_{\scriptscriptstyle S})
\end{array}
\Biggl|\Biggr|
\begin{array}{c}
[f_{\scriptscriptstyle CS}]\\
([2^2]_{\scriptscriptstyle CS}{\times}[2]_{\scriptscriptstyle CS})
\end{array}\right),&\nonumber\\
&\Gamma_{\scriptscriptstyle C\cdot S}^{S\!=\!1}
([f_{\scriptscriptstyle CS}]([21^2]_{\scriptscriptstyle CS}
{\times}[1^2]_{\scriptscriptstyle CS}),s_{56}\!=\!1)
\equiv\left(\begin{array}{cc}
[2^3]_{\scriptscriptstyle C}&[42]_{\scriptscriptstyle S}\\
([21^2]_{\scriptscriptstyle C}{\times}[1^2]_{\scriptscriptstyle C})&
([2^2]_{\scriptscriptstyle S}{\times}[2]_{\scriptscriptstyle S})
\end{array}
\Biggl|\Biggr|
\begin{array}{c}
[f_{\scriptscriptstyle CS}]\\
([21^2]_{\scriptscriptstyle CS}{\times}[1^2]_{\scriptscriptstyle CS})
\end{array}\right),&\nonumber\\
&\Gamma_{\scriptscriptstyle C\cdot S}^{S\!=\!1}
([f_{\scriptscriptstyle CS}]([21^2]_{\scriptscriptstyle CS}
{\times}[1^2]_{\scriptscriptstyle CS}),s_{56}\!=\!0)
\equiv\left(\begin{array}{cc}
[2^3]_{\scriptscriptstyle C}&[42]_{\scriptscriptstyle S}\\
([21^2]_{\scriptscriptstyle C}{\times}[1^2]_{\scriptscriptstyle C})&
([31]_{\scriptscriptstyle S}
{\times}[2]_{\scriptscriptstyle S})
\end{array}
\Biggl|\Biggr|
\begin{array}{c}
[f_{\scriptscriptstyle CS}]\\
([21^2]_{\scriptscriptstyle CS}{\times}[1^2]_{\scriptscriptstyle CS})
\end{array}\right),&
\label{ap9}
\end{eqnarray}
Here the $[f_{\scriptscriptstyle CS}]$ are all the $CS$-Young schemes from the inner product
\begin{eqnarray}
[f_{\scriptscriptstyle CS}]=[2^3]_{\scriptscriptstyle C}\circ
[42]_{\scriptscriptstyle S}=[42],\quad[321],\quad[2^3],\quad[31^3],
\quad[21^4]
\label{ap10}
\end{eqnarray}
Values of all the necessary scalar factors~(\ref{ap9}) are shown in 
Tables~\ref{table2} and \ref{table3}.

Only the Young schemes $[f_{\scriptscriptstyle CST}]=[2^21^2],\,\,
[21^4],\,\,[1^6]$ are important for configurations $d_f$, $d^{\prime\prime}$
and $d_0$ ($[f_{\scriptscriptstyle CST}]=[\tilde f_{\scriptscriptstyle X}]$,
where $[\tilde f_{\scriptscriptstyle X}]$ is the Young scheme conjugated to
$[f_{\scriptscriptstyle X}]$). All the necessary scalar factors
\begin{eqnarray}
&\Gamma_{\scriptscriptstyle CS\cdot T}^{T\!=\!0}([f_{\scriptscriptstyle CST}]:
[f_{\scriptscriptstyle CS}]([2^2]_{\scriptscriptstyle CS}
{\times}[2]_{\scriptscriptstyle CS}),t_{56}\!=\!0)
\qquad\qquad\qquad\qquad\qquad\qquad\qquad\qquad&\nonumber\\
&\equiv\left(\begin{array}{cc}
[f_{\scriptscriptstyle CS}]&[3^2]_{\scriptscriptstyle T}\\
([2^2]_{\scriptscriptstyle CS}{\times}[2]_{\scriptscriptstyle CS})&
([2^2]_{\scriptscriptstyle T}
{\times}[1^2]_{\scriptscriptstyle T})
\end{array}
\Biggl|\Biggr|
\begin{array}{c}
[f_{\scriptscriptstyle CST}]\\
([1^4]_{\scriptscriptstyle CST}{\times}[1^2]_{\scriptscriptstyle CST})
\end{array}\right),&\nonumber\\
&\Gamma_{\scriptscriptstyle CS\cdot T}^{T\!=\!0}([f_{\scriptscriptstyle CST}]:
[f_{\scriptscriptstyle CS}]([21^2]_{\scriptscriptstyle CS}
{\times}[1^2]_{\scriptscriptstyle CS}),t_{56}\!=\!1)
\qquad\qquad\qquad\qquad\qquad\qquad\qquad\qquad&\nonumber\\
&\equiv\left(\begin{array}{cc}
[f_{\scriptscriptstyle CS}]&[3^2]_{\scriptscriptstyle T}\\
([21^2]_{\scriptscriptstyle CS}{\times}[2]_{\scriptscriptstyle CS})&
([31]_{\scriptscriptstyle T}
{\times}[2]_{\scriptscriptstyle T})
\end{array}
\Biggl|\Biggr|
\begin{array}{c}
[f_{\scriptscriptstyle CST}]\\
([1^4]_{\scriptscriptstyle CST}{\times}[1^2]_{\scriptscriptstyle CST})
\end{array}\right)&
\label{ap11}
\end{eqnarray}
are shown in Table~\ref{table4}. The coefficients 
$\Gamma_{\scriptscriptstyle X\cdot CST}$ are trivial weight factors
$\Gamma_{\scriptscriptstyle X\cdot CST}
([6]_{\scriptscriptstyle X}([4]{\times}[2]))=1$,
$\Gamma_{\scriptscriptstyle X\cdot CST}
([51]_{\scriptscriptstyle X}([4]{\times}[2]))=\sqrt{\frac{1}{5}}$ and
$\Gamma_{\scriptscriptstyle X\cdot CST}
([42]_{\scriptscriptstyle X}([4]{\times}[2]))=\sqrt{\frac{1}{9}}$
dependent only on the dimensions of irredusible representations of the
symmetrical group for given Young schemes: $n_{[6]}=1$, $n_{[51]}=5$ and
$n_{[42]}=9$. The last factor in the right-hand side of Eq.~(\ref{ap8}),
the orbital f.p.c. of TISM $\Gamma_{\scriptscriptstyle X}$, takes the value
dependent on the configuration, i.e. only five different values of 
$\Gamma_{\scriptscriptstyle X}$ are necessary: 
\begin{eqnarray}
&\Gamma_{\scriptscriptstyle X}(s^6[6](s^4[4]{\times} s^2[2]))=1,&\nonumber\\
&\Gamma_{\scriptscriptstyle X}(s^4p^2-s^52s[6](s^4[4]{\times} s2s[2]))=
\sqrt{\frac{1}{5}},&\nonumber\\
&\Gamma_{\scriptscriptstyle X}(s^4p^2-s^52d[6](s^4[4]{\times} s2d[2]))=
\sqrt{\frac{1}{5}},&\nonumber\\
&\Gamma_{\scriptscriptstyle X}(s^5p[51](s^4[4]{\times} sp[2]))=
-\sqrt{\frac{3}{5}},&\nonumber\\
&\Gamma_{\scriptscriptstyle X}(s^4p^2[42](s^4[4]{\times} p^2[2])L=0,2)=
-\sqrt{\frac{3}{10}}.&
\label{ap12}
\end{eqnarray}

Thus the total transition amplitude (\ref{ap6}) are expressed through the
product of factors (\ref{ap1})--(\ref{ap2}), (\ref{ap9})--(\ref{ap12}) summed
over states of the pair "56" (the summation should be extended over all the
possible two-quark states, but the fixed quantum numbers of the initial,
intermediate and final states impose the restriction that only the summation 
over $j_{56}=0,\,1$ and 
$[f_{\scriptscriptstyle 56}]_{\scriptscriptstyle CS}=[2],\,[1^2]$ is allowed):
\begin{eqnarray}
&15\,\langle d_0|H^{(5)}_{\pi qq}(k_5)|d^{\prime\prime}\rangle\,
\langle d^{\prime\prime}|H^{(6)}_{\pi qq}(k_6)|d_f\rangle =
15\sum\limits_{j_{56}=0,\,1}
\sum\limits_{[f_{\scriptscriptstyle 56}]_{\scriptscriptstyle CS}=[2],[1^2]}
\Gamma_{\scriptscriptstyle X\cdot CST}
([6]_{\scriptscriptstyle X}([4]{\times}[2]))&\nonumber\\
&{\times}\left[\Gamma_{\scriptscriptstyle X\cdot CST}
([51]_{\scriptscriptstyle X}([4]{\times}[2]))\right]^2\,
\Gamma_{\scriptscriptstyle X\cdot CST}
([42]_{\scriptscriptstyle X}([4]{\times}[2]))\,\,
\Gamma_{\scriptscriptstyle X}(s^6[6](s^4[4]{\times} s^2[2]))&\nonumber\\
&{\times}\left[\Gamma_{\scriptscriptstyle X}(s^5p[51](s^4[4]{\times} sp[2]))
\right]^2\,\,
\Gamma_{\scriptscriptstyle X}(s^4p^2[42](s^4[4]{\times} 
p^2[2])L=0,2)&\nonumber\\
&{\times}\Gamma_{\scriptscriptstyle C\cdot S}^{S\!=\!1}([2^3]_{\scriptscriptstyle CS}
([f_{\scriptscriptstyle 1234}]_{\scriptscriptstyle CS}
{\times}[f_{\scriptscriptstyle 56}]_{\scriptscriptstyle CS}),
s_{\scriptscriptstyle 56})\,\,
\left[\Gamma_{\scriptscriptstyle C\cdot S}^{S\!=\!0}
([2^21^2]_{\scriptscriptstyle CS}
([f_{\scriptscriptstyle 1234}]_{\scriptscriptstyle CS}
{\times}[f_{\scriptscriptstyle 56}^{\prime}]_{\scriptscriptstyle CS}),
s_{\scriptscriptstyle 56}^{\prime})\right]^2&\nonumber\\
&{\times}\Gamma_{\scriptscriptstyle C\cdot S}^{S\!=\!1}
([f_{\scriptscriptstyle CS}]
([f_{\scriptscriptstyle 1234}]_{\scriptscriptstyle CS}
{\times}[f_{\scriptscriptstyle 56}]_{\scriptscriptstyle CS}),
s_{\scriptscriptstyle 56})\,\,
\Gamma_{\scriptscriptstyle CS\cdot T}^{T\!=\!0}
([1^6]_{\scriptscriptstyle CST}\!\!:[2^3]_{\scriptscriptstyle CS}
([f_{\scriptscriptstyle 1234}]_{\scriptscriptstyle CS}
{\times}[f_{\scriptscriptstyle 56}]_{\scriptscriptstyle CS}),t_{56})
&\nonumber\\
&{\times}\left[\Gamma_{\scriptscriptstyle CS\cdot T}^{T\!=\!1}
([21^4]_{\scriptscriptstyle CST}\!\!:[2^21^2]_{\scriptscriptstyle CS}
([f_{\scriptscriptstyle 1234}]_{\scriptscriptstyle CS}
{\times}[f_{\scriptscriptstyle 56}^{\prime}]_{\scriptscriptstyle CS}),
t_{56}^{\prime})\right]^2&\nonumber\\
&{\times}\Gamma_{\scriptscriptstyle CS\cdot T}^{T\!=\!0}
([2^21^2]_{\scriptscriptstyle CST}\!\!:[f_{\scriptscriptstyle CS}]
([f_{\scriptscriptstyle 1234}]_{\scriptscriptstyle CS}
{\times}[f_{\scriptscriptstyle 56}]_{\scriptscriptstyle CS}),t_{56})\,\,
T^{(5)}_{1p\to 0s}(j_{56})\,\,T^{(6)}_{2s(2d)\to 1p}(j_{56})&
\label{ap13}
\end{eqnarray}

Spin and isospin of the quark pair 
$s_{\scriptscriptstyle 56}(s_{\scriptscriptstyle 56}^{\prime})$, 
$t_{\scriptscriptstyle 56} (t_{\scriptscriptstyle 56}^{\prime})$ in 
Eq.~(\ref{ap13}) depend on color quntum numbers of the pair.  
For example, $t_{\scriptscriptstyle 56}=1 
(t^{\prime}_{\scriptscriptstyle 56}=0$) for 
$[f_{\scriptscriptstyle 56}]_{\scriptscriptstyle CS}=[1^2]$ and
$t_{\scriptscriptstyle 56}=0
(t^{\prime}_{\scriptscriptstyle 56}=1$) for 
$[f_{\scriptscriptstyle 56}]_{\scriptscriptstyle CS}=[2]$. A general
rule for $s_{\scriptscriptstyle 56}(s_{\scriptscriptstyle 56}^{\prime})$
is easy to understand from the right-hand side of Eqs.~(\ref{ap9}). One
can remark
that in the case of L=2 (the $^3D_1$ initial state) the value 
$j_{\scriptscriptstyle 56}=0$ does not contribute to the transition
$^3D_1\to^3S_1$ and only the term with $j_{\scriptscriptstyle 56}=1$
should be taken in the right-hand side of Eq.~(\ref{ap13}). This leads
to a difference in the value of the coupling constant
$f_{\pi d^{\prime\prime}d_f}$ for L=0 and 2 that is marked by an additional 
superscript L: $f^L_{\pi d^{\prime\prime}d_f}$. 

Calculated values of the product
$f^L_{\pi d^{\prime\prime}d_f}\,f_{\pi d_0d^{\prime\prime}}$ are
shown in Table~\ref{table5}.    
Substituting these values into Eq.~(\ref{gL}) one obtains the following
expression for factors $g_{\scriptscriptstyle L}$ in the transition 
operator~(\ref{om1}) 
\begin{eqnarray}
g_{\scriptscriptstyle L}=g_{\pi\pi\sigma}\frac{f_{\pi qq}^2}{m_{\pi}^2}\,
\frac{1}{m_q^2b^2}\,\frac{1}{180}\times
\left\{\begin{array}{cr}-\frac{617}{1620\sqrt{5}},&L=0\\
\frac{55}{324},&L=2
\end{array}\right.
\label{ap14}
\end{eqnarray}

\newpage 

\begin{table}
\caption{The model parameters for the different partial waves.}
\label{table1}
\begin{tabular}{cccccccccc}\\ 
$^{2s+1}L_J$& $\Lambda$ (GeV) &$r_0^{\rm orth}$ (fm) & $\lambda_{11}$& $\lambda_{22}$&
$\lambda_{12}$& $r_{0_1}$ (fm) &$r_{0_2}$ (fm) & $E_0$ (MeV) & $\chi^2$ \\
\hline
$^1S_0(< 600$ MeV)& 0.65 & 0.3943 & 2.055 &&& 0.59686 && 356 & 1.09 \\ \hline
$^1S_0(< 1.2$ GeV)& 0.65 & 0.3943 & 4.565 &&& 0.5106 && 550 & 3.9 \\ \hline
$^1D_2$           & 0.65 &        & 0.02463 &&& 0.79403 && 330 & 0.028\\ \hline
$^3S_1-{ }^3D$         & 0.5936 & 0.3737 & 7.201 &0.007928 & 0.2294 
                        &0.45469 &0.65652 & 681 & 1.7 \\ \hline
$^3D_2$           & 0.5527 &    & 0.01038   &&& 0.86037&& 800 & 0.062 \\ \hline
$^3D_3-{ }^3G_3$  & 0.5936 &    & 0.002927 &0.1753 & 0.02624 & 0.89971 
                               & 0.42893 & 800 & 0.11 \\ \hline
$^1P_1$           & 0.7324 & 0.46572 & 28.74  &&& 0.44311 && 600 & 0.167 \\ \hline
$^3P_0$           & 0.65 & 0.3445 & 0.02841 &&& 0.455  && 400 & 0.14 \\ \hline
$^3P_1$           & 0.65 & 0.4491 & 3.195 &&& 0.51749 && 600 & 0.13 \\ \hline
$^3P_2-{ }^3F_2$  & 0.65 &      & 0.03124 & -0.006486 & 0.000765 & 0.70995  
                               & 0.75653 & 360 & 0.71\\  
\end{tabular}
\end{table}

\begin{table}
\caption{Scalar factors $\Gamma^{\scriptscriptstyle S=1(0)}_{\scriptscriptstyle 
C\cdot S}([f_{\scriptscriptstyle CS}]
([f_{\scriptscriptstyle CS}^{\prime}]\!{\times}\!
[f_{\scriptscriptstyle CS}^{\prime\prime}]),
s_{\scriptscriptstyle 56 })$  of the Clebsch-Gordon 
coefficients for the group
$SU(6)_{\scriptscriptstyle CS}\subset SU(3)_{\scriptscriptstyle C}
{\times} SU(2)_{\scriptscriptstyle S}$ (see Eq.~(\protect\ref{ap9})).}
\label{table2}
\begin{tabular}{lcccc}
&\multicolumn{4}{c}{$S=1$}\\[5pt]
&\multicolumn{2}{c}{$[2^2]_{\scriptscriptstyle CS}%
{\times}[2]_{\scriptscriptstyle CS}$}&%
\multicolumn{2}{c}{$[21^2]_{\scriptscriptstyle CS}
{\times}[1^2]_{\scriptscriptstyle CS}$}\\[5pt]
\tableline
&$[2^2]_{\scriptscriptstyle C}{\times}[2]_{\scriptscriptstyle C}$&%
$[21^2]_{\scriptscriptstyle C}{\times}[1^2]_{\scriptscriptstyle C}$&%
$[2^2]_{\scriptscriptstyle C}{\times}[2]_{\scriptscriptstyle C}$%
&$[21^2]_{\scriptscriptstyle C}{\times}[1^2]_{\scriptscriptstyle C}$\\[5pt]
&$[2^2]_{\scriptscriptstyle S}{\times}[2]_{\scriptscriptstyle S}$%
&$[31]_{\scriptscriptstyle S}{\times}[1^2]_{\scriptscriptstyle S}$%
&$[31]_{\scriptscriptstyle S}{\times}[1^2]_{\scriptscriptstyle S}$%
&$[2^2]_{\scriptscriptstyle S}{\times}[2]_{\scriptscriptstyle S}$\\[5pt]
\tableline
$[42]_{\scriptscriptstyle CS}:$&$\sqrt{\frac{1}{20}}$&$-\sqrt{\frac{9}{20}}$%
&$0$&$0$\\[5pt]
$[321]_{\scriptscriptstyle CS}:$&$\sqrt{\frac{8}{15}}$%
&$-\sqrt{\frac{2}{15}}$&$\sqrt{\frac{2}{9}}$%
&$\sqrt{\frac{8}{27}}$\\[5pt]
$[2^3]_{\scriptscriptstyle CS}:$&$\sqrt{\frac{5}{12}}$%
&$\sqrt{\frac{5}{12}}$&$\sqrt{\frac{5}{18}}$%
&$-\sqrt{\frac{5}{54}}$\\[5pt]
$[31^3]_{\scriptscriptstyle CS}:$&$0$%
&$0$&$-\sqrt{\frac{1}{18}}$%
&$-\sqrt{\frac{25}{54}}$\\[5pt]
$[21^4]_{\scriptscriptstyle CS}:$&$0$&$0$&$\sqrt{\frac{4}{9}}$%
&$-\sqrt{\frac{4}{27}}$\\[5pt]
\end{tabular}
\end{table}

\begin{table}
\caption{Scalar factors $\Gamma^{\scriptscriptstyle S=1(0)}_{\scriptscriptstyle 
C\cdot S}([f_{\scriptscriptstyle CS}]
([f_{\scriptscriptstyle CS}^{\prime}]\!{\times}\!
[f_{\scriptscriptstyle CS}^{\prime\prime}]),
s_{\scriptscriptstyle 56 })$  of the Clebsch-Gordon 
coefficients for the group
$SU(6)_{\scriptscriptstyle CS}\subset SU(3)_{\scriptscriptstyle C}
{\times} SU(2)_{\scriptscriptstyle S}$ (continued).}
\label{table3}
\begin{tabular}{lcccc}
&\multicolumn{4}{c}{$S=0$}\\[5pt]
&\multicolumn{2}{c}{$[2^2]_{\scriptscriptstyle CS}
{\times}[1^2]_{\scriptscriptstyle CS}$}&%
\multicolumn{2}{c}{$[21^2]_{\scriptscriptstyle CS}
{\times}[2]_{\scriptscriptstyle CS}$}\\[5pt]
\tableline
&$[2^2]_{\scriptscriptstyle C}{\times}[2]_{\scriptscriptstyle C}$%
&$[21^2]_{\scriptscriptstyle C}{\times}[1^2]_{\scriptscriptstyle C}$%
&$[2^2]_{\scriptscriptstyle C}{\times}[2]_{\scriptscriptstyle C}$%
&$[21^2]_{\scriptscriptstyle C}{\times}[1^2]_{\scriptscriptstyle C}$\\[5pt]
&$[2^2]_{\scriptscriptstyle S}{\times}[1^2]_{\scriptscriptstyle S}$%
&$[31]_{\scriptscriptstyle S}{\times}[2]_{\scriptscriptstyle S}$&%
$[31]_{\scriptscriptstyle S}{\times}[2]_{\scriptscriptstyle S}$%
&$[2^2]_{\scriptscriptstyle S}{\times}[1^2]_{\scriptscriptstyle S}$\\[5pt]
\tableline
$[2^21^2]_{\scriptscriptstyle CS}:$&%
$-\sqrt{\frac{3}{4}}$&$\sqrt{\frac{1}{4}}$&$-\sqrt{\frac{1}{2}}$%
&$\sqrt{\frac{1}{2}}$\\[5pt]
\end{tabular}
\end{table}

\begin{table}
\caption{Scalar factors 
$\Gamma^{\scriptscriptstyle T=0(1)}_{\scriptscriptstyle 
CS\cdot T}([f_{\scriptscriptstyle CST}]:[f_{\scriptscriptstyle CS}]
([f_{\scriptscriptstyle CS}^{\prime}]\!{\times}\!
[f_{\scriptscriptstyle CS}^{\prime\prime}]),
t_{\scriptscriptstyle 56 })$  of the Clebsch-Gordon 
coefficients for the group
$SU(12)_{\scriptscriptstyle CST}\subset SU(6)_{\scriptscriptstyle CS}
{\times} SU(2)_{\scriptscriptstyle T}$ (see Eq.~(\protect\ref{ap11})).}
\label{table4}
\begin{tabular}{ccccccc}
&\multicolumn{5}{c}{$T=0$}&\multicolumn{1}{c}{$T=1$}\\[5pt]
&$[42]_{\scriptscriptstyle CS}$&$[321]_{\scriptscriptstyle CS}$%
&$[2^3]_{\scriptscriptstyle CS}$&$[31^3]_{\scriptscriptstyle CS}$%
&$[21^4]_{\scriptscriptstyle CS}$&$[2^21^2]_{\scriptscriptstyle CS}$\\[5pt]
\tableline
$[2^21^2]_{\scriptscriptstyle CST}(T=0):$&&&&&&\\
$([2^2]_{\scriptscriptstyle CS}{\times}[2]_{\scriptscriptstyle CS})\circ%
([2^2]_{\scriptscriptstyle T}{\times}[1^2]_{\scriptscriptstyle 
T})$&$1$&$-\sqrt{\frac{3}{8}}$&$-\sqrt{\frac{3}{5}}$&$0$&$0$&--\\[5pt]
$([21^2]_{\scriptscriptstyle CS}{\times}[1^2]_{\scriptscriptstyle CS})\circ%
([31]_{\scriptscriptstyle T}{\times}[2]_{\scriptscriptstyle 
T})$&$0$&$\sqrt{\frac{5}{8}}$&$-\sqrt{\frac{2}{5}}$&$1$&$1$&--\\[5pt]
\tableline
$[1^6]_{\scriptscriptstyle CST}(T=0):$&&&&&&\\
$([2^2]_{\scriptscriptstyle CS}{\times}[2]_{\scriptscriptstyle CS})\circ%
([2^2]_{\scriptscriptstyle T}{\times}[1^2]_{\scriptscriptstyle T})$%
&$0$&$0$&$-\sqrt{\frac{2}{5}}$&$0$&$0$&--\\[5pt]
$([21^2]_{\scriptscriptstyle CS}{\times}[1^2]_{\scriptscriptstyle CS})\circ%
([31]_{\scriptscriptstyle T}{\times}[2]_{\scriptscriptstyle 
T})$&$0$&$0$&$\sqrt{\frac{3}{5}}$&$0$&$0$&--\\[5pt]
\tableline
$[2^21^2]_{\scriptscriptstyle CST}(T=1):$&&&&&&\\
$([2^2]_{\scriptscriptstyle CS}{\times}[2]_{\scriptscriptstyle CS})\circ%
([2^2]_{\scriptscriptstyle T}{\times}[1^2]_{\scriptscriptstyle T})$%
&--&--&--&--&--&$-\sqrt{\frac{4}{9}}$\\[5pt]
$([21^2]_{\scriptscriptstyle CS}{\times}[1^2]_{\scriptscriptstyle CS})\circ%
([31]_{\scriptscriptstyle T}{\times}[2]_{\scriptscriptstyle T})$%
&--&--&--&--&--&$\sqrt{\frac{1}{6}}$\\[5pt]
\end{tabular}
\end{table}

\begin{table}
\caption{The products of coupling constants $f_{\pi d_0d^{\prime\prime}}\,
f^{\scriptscriptstyle L=0(2)}_{\pi d^{\prime\prime}d_f}$ for the two-step transition
$d_f\to d^{\prime\prime}+\pi\to d_0+\pi+\pi$ with creation of the
scalar-isoscalar $\pi+\pi$ pair ("$\sigma$-meson") and the overlap factor
between the $NN$ and $d_f$ states $U^{NN}_f$.}
\label{table5}
\begin{tabular}{ccccccc}
&\multicolumn{6}{c}{Quantum numbers of $d_f$}\\[5pt]
&\multicolumn{1}{c}{$(s^4p^2\!-\!s^52s(2d))[6]_{\scriptscriptstyle X}$}%
&\multicolumn{5}{c}{$s^4p^2[42]_{\scriptscriptstyle X}L=0(2)$}\\[5pt]
&$[2^3]_{\scriptscriptstyle CS}$%
&$[42]_{\scriptscriptstyle CS}$&$[321]_{\scriptscriptstyle CS}$%
&$[2^3]_{\scriptscriptstyle CS}$&$[31^3]_{\scriptscriptstyle CS}$%
&$[21^4]_{\scriptscriptstyle CS}$\\[5pt]
\tableline
$180{\times} f_{\pi d_0d^{\prime\prime}}\,
f^{\scriptscriptstyle L}_{\pi d^{\prime\prime}d_f}:$&&&&&&\\
$L=0$&$\frac{17}{18}$&$-\sqrt{\frac{1}{5}}$%
&$-\frac{71}{72}\sqrt{\frac{1}{5}}$&$-\frac{13}{18}$%
&$\frac{1}{9\sqrt{2}}$&$-\frac{5}{9}$\\[5pt]
$L=2$&$\frac{13\sqrt{5}}{36}$&$\frac{1}{2}$&$-\frac{31}{36}$%
&$-\frac{17\sqrt{5}}{36}$&$-\frac{5\sqrt{5}}{36\sqrt{2}}$%
&$-\frac{\sqrt{5}}{18}$\\[5pt]
\tableline
$U^{NN}_f:$%
&$\sqrt{\frac{1}{9}}$&$-\sqrt{\frac{9}{20}}$&$\sqrt{\frac{16}{45}}$%
&$\sqrt{\frac{1}{36}}$&$-\sqrt{\frac{1}{18}}$&$0$\\[5pt]
\end{tabular}
\end{table}
 
\centerline{\large \bf Figure captions} 
 
\bigskip 
 
{\bf FIG.1.}  The traditional t-channel meson-exchange 
mechanism (a) compared to the new $s$-channel "dressed" bag mechanism (b) 
for $NN$ interaction.
\bigskip 
 
{\bf FIG.2.}  Schematic representation of the 
two-pion emission in the transition of two $p$-shell quarks to an $s$
orbit.
\bigskip 

{\bf FIG.3.} The graph illustrating the $\sigma$- (or 
$\rho$-) meson emission and subsequent absorption by diquark pairs in  
the intermediate six-quark bag-like state.
\bigskip 

{\bf FIG.4.} The graph illustrates two sequential 
$\pi$-meson emissions and absorptions via an intermediate $\sigma$- (or 
$\rho$-) meson and the generation of a six-quark bag.
 
\bigskip 
{\bf FIG.5.} The kinematic variables in the 
triangle diagram corresponding to the $\sigma$- (or $\rho$-) meson 
generation from two $\pi$-mesons emerging in the transition of two 
$p$-shell quarks to the $s$-orbit (see also FIG.~4). 
 
\bigskip 
{\bf FIG.6.}  The $NN$ phase shifts (in deg.) in our 
model in comparison with PSA data (SAID, solution SP99). 
 
\bigskip 
{\bf FIG.6.} ({\em Continued.}) 

\bigskip 
{\bf FIG.7.} The mixing parameter $\varepsilon_1$ for different values 
of cut-off parameter $\Lambda_{\pi NN}$ (see text). 

\bigskip 
{\bf FIG.8.} The $^1S_0$ phase shifts fitted by means of our 
model (\ref{mod1}) until $E_{\rm lab}=1200$~MeV.

\bigskip  {\bf FIG.9.} Some graphs illustrating the new-type of $3N$  forces. 

\end{document}